\documentclass[aps,pre,showpacs,superscriptaddress]{revtex4} 
\usepackage[T2A]{fontenc}
\usepackage[cp1251]{inputenc}
\usepackage[russian,english]{babel}
\usepackage{graphicx}
\usepackage{color}
\usepackage{bm}
\usepackage{amssymb}
\usepackage[intlimits]{amsmath}
\usepackage{amsmath}
\usepackage{epsfig}
\usepackage{subfig}
\DeclareGraphicsExtensions{.pdf,.eps,.jpg}

\newcommand{\be}{\begin{equation}}
\newcommand{\ee}{\end{equation}}
\newcommand{\ben}{\begin{equation*}}
\newcommand{\een}{\end{equation*}}

\begin{document}
\title{Optimal input signal distribution for nonlinear optical fiber channel with small Kerr nonlinearity.}


\begin{abstract}
We consider the information channel described by Schr\"{o}dinger
equation with  additive Gaussian noise. We introduce the model of
the input signal and the model of the output signal receiver. For
this channel, using perturbation theory for the small nonlinearity
parameter, we calculate the first three terms of the expansion of
the conditional probability density function in the nonlinearity parameter. At
large signal-to-noise power ratio  we calculate the
conditional entropy, the output signal entropy, and the mutual
information in the leading and next-to-leading order in the
nonlinearity parameter and in the leading order in the parameter
$1/\mathrm{SNR}$. Using the mutual information we find the optimal
input signal distribution and channel capacity in the leading and
next-to-leading order in the nonlinearity parameter. Finally, we present the method of the construction of the input signal with the optimal statistics for the given shape of the signal.

\end{abstract}

\hskip 2cm
\date{\today}

\author{A.~V.~Reznichenko}
\email[E-mail: ]{A.V.Reznichenko@inp.nsk.su} \affiliation{Budker
Institute of  Nuclear Physics of Siberian Branch Russian Academy of
Sciences, Novosibirsk, 630090 Russia} \affiliation{Novosibirsk State
University, Novosibirsk, 630090 Russia}

\author{A.~I.~Chernykh}
\email[Electronic address:]{chernykh@iae.nsk.su}
\affiliation{Institute of Automation and Electrometry of the
Siberian Branch of the Russian Academy of Sciences, Novosibirsk,
630090 Russia} \affiliation{Novosibirsk State University,
Novosibirsk, 630090 Russia}

\author{E.~V.~Sedov}
\email[Electronic address:]{e.sedov@g.nsu.ru}
\affiliation{Novosibirsk State University, Novosibirsk, 630090
Russia} \affiliation{Aston Institute of Photonics Technologies,
Aston University, Aston  Triangle, Birmingham, B4 7ET, UK}

\author{I.~S.~Terekhov}
\email[E-mail: ]{I.S.Terekhov@gmail.com} \affiliation{Budker
Institute of  Nuclear Physics of Siberian Branch Russian Academy of
Sciences, Novosibirsk, 630090 Russia} \affiliation{Novosibirsk State
University, Novosibirsk, 630090 Russia}


\maketitle

\section{ Introduction}

Nowadays the fiber-optic communication channels are actively
developed. That is why it is important to know their maximum
information transmission rate, i.e., the channel capacity. For small
powers of the outcoming signal, these channels are well described by
the linear models. For the powers in question, the capacity of the
noisy channel was found analytically in Shannon's famous papers
\cite{Shannon:1948,Shannon:1949}:
\begin{eqnarray}\label{CapacityShannon}
C\propto\log\left(1+\mathrm{SNR}\right)\,,
\end{eqnarray}
where $\mathrm{SNR}=P/N$ is the signal-to-noise power ratio, $P$ is
the average input signal power, and $N$ is the noise power. As one
can see from this relation, to increase the  channel capacity with
the noise power being fixed it is necessary to increase the signal
power. However, as the signal power increases, the nonlinear effects
become more important.  In this case, a simple expression for the
capacity of this noisy nonlinear channel is unknown. The reason is
that the expression should take into account all the details of the
mathematical model of the nonlinear channel. This model implies the
following components: the input signal model, i.e., the method for
encoding incoming information, the physical signal propagation model
across the fiber wire, the receiver model (i.e., the signal
detection features, frequency filtering), and the procedure of the
signal post-processing. Thus, for different channel
models the expressions for the capacity will also be different even
for some matching components. It is very difficult to find an
explicit expression for the capacity even for a specific (and often
highly simplified) model  of a nonlinear communication channel. A
more realistic problem for such channels is to find the upper or
lower bounds for the capacity. For example, in the case of commonly
used models involving the signal propagation governed by the
nonlinear Schr\"odinger equation (NLSE) with additive white Gaussian
noise, see Refs.
\cite{Haus:1991,Mecozzi:1994,Iannoe:1998,Turitsyn:2000} and
references therein, an expression for the capacity has not yet been
obtained. However, in the case when a small parameter is present in
the model, often it is possible to invoke the perturbation theory
for this parameter if the zero-parameter model turns out to be
solvable. For instance, in the case of the low noise power in the
channel (i.e., for the large $\mathrm{SNR}$ parameter), one can
develop an analog of the semiclassical approach in quantum mechanics
\cite{Terekhov:2014}. Further, in the case when the coefficient of
the second  dispersion in the channel is small, it is also possible
to construct the perturbation theory based on this parameter
\cite{ReznichenkoIEEE:2018,ReznichenkoIOP:2018,
ReznichenkoCor:2018}, since the zero-parameter model is well
developed:
\cite{M1994,MS:2001,Tang:2001,tdyt03,Mansoor:2011,Terekhov:2016a,Terekhov:2017,
Kramer:2017,Terekhov:2018a}. Finally, in the case of the moderate
input power, it is also possible to develop the perturbation theory
for the Kerr nonlinearity parameter of the fiber: see
\cite{Kramer:2020}. In different approaches, the
capacity for an optical fiber channel with nonzero dispersion and Kerr nonlinearity has
been studied both analytically and numerically, see Refs.
\cite{Mitra:2001,Narimanov:2002,Kahn:2004,Essiambre:2008,Essiambre:2010,
Killey:2011,Agrell:2014, Sorokina2014} and references therein.

In the present paper, we focus on the study of the nonlinear channel
described by NLSE with additive Gaussian noise using the
perturbation theory in the small parameter of Kerr nonlinearity and
large   $\mathrm{SNR}$. To this end, we
consistently build the model of the input signal $X(t)$, we study
the impact of the spectral noise width on the output signal $Y(t)$
(i.e. the raw signal in the receiving point), and we investigate the
influence of the signal detection procedure in the receiver and the
post-processing, i.e., the procedure of the input data extraction
from the received signal $Y(t)$. We carry out all our calculations
in the leading and next-to-leading orders in the Kerr nonlinearity
parameter.

To study the mutual information we use the representation for the
conditional probability density function $P[Y (t)|X(t)]$, i.e. the
probability density to get the output signal $Y(t)$, if the input
signal is $X (t)$, through the path-integral \cite{Terekhov:2014}.
This representation for $P[Y (t)|X(t)]$ is especially convenient to use the perturbation theory. Generally speaking, the function
spaces of the input signal $X(t)$ and the output signal $Y(t)$ are
infinite-dimensional. However, information is transmitted using some
finite set pulses of a certain shape, spread either in time or in
frequency space. For example, the input signal $X(t)$ can be constructed as
follows
\begin{eqnarray}
X(t)=\sum^{M}_{k=-M} C_{k} \, s(t-k T_0).
\end{eqnarray}
Here $s(t)$ is a fixed envelope function of time, $C_k$ --- are
complex variables that carry information about the input signal,
$T_0$ is a time interval between two successive pulses. The
problem of information transfer is reduced to recover the
coefficients $\{C_{- M}, \ldots, C_{M} \}$ from the signal $Y(t)$
received at the output. To find the informational characteristics
of the communication channel, we need to reduce the density
functional $P[Y(t)|X(t)]$ to the functional $P[\{\tilde {C}\}|\{C\}]
$, i.e., the conditional probability density to get a set of coefficients
$\{\tilde{C}_k \}$, if the input signal was encrypted by the
coefficients $\{C_k\}$. The functional $ P[\{\tilde{C}\}|\{C\}]$
depends both on the physical laws of  the signal propagation along
the communication channel, and on the detection procedure with the
post-processing of the signal. Generally, the functional
$P[\{\tilde{C} \}|\{C\}]$ can not be reduced to a factorized form
\begin{eqnarray}\label{factor}
P[\{\tilde{C}\}|\{C\}]= \prod^{M}_{k=-M} P^{(k)}[\tilde{C}_k|C_k]
\end{eqnarray}
due to the dispersion effects in the first place. This means that we
deal with a communication channel with memory (commonly, with
infinite one). Fiber optical channels with memory were previously
considered in a bulk of papers, see for example
\cite{AgrelKarlson:2014}.

In our work we calculate the density $ P[\{\tilde{C}\}|\{C\}]$ for
nonlinear fiber optic communication channel, in which the signal
propagation is governed by the nonlinear  Schr\"odinger equation
with additive Gaussian noise of finite spectral width. Our model
also includes a receiver model and post-processing procedure of the
extraction of the coefficients $\{\tilde{C}_k\}$ from the detected
signal $Y(t)$. The density functional $ P[Y(t)|X(t)]$, as well as
the density $ P[\{\tilde{C}\}|\{C \}]$ were found with the use of
two different methods. The first method is based on the direct
calculation of the path-integral representing $P[Y(t)|X(t)]$ via the
effective two-dimensional action \cite{Terekhov:2014} in the leading
and next-to-leading orders in the parameter $1/\mathrm{SNR}$ and in
the parameter of Kerr nonlinearity, correspondingly. The second method is based on the independent calculation of the correlators of the solution of the
nonlinear Schr\"{o}dinger equation with additive Gaussian
noise for a fixed input signal $X(t)$.

Using the found density $P[\{\tilde{C}\}|\{C\}]$ we calculated the
entropy of the output signal and the conditional entropy. It allowed
us to find mutual information in a leading and next-to-leading
orders in the parameter $1/\mathrm{SNR}$ and in the parameter of the Kerr
nonlinearity. Then we found the extremum of the mutual information
and calculated the probability density of the input signal $P_{opt}[\{C \}]$
that delivers this extremum. We demonstrated that in the first
non-vanishing order in the Kerr nonlinearity, the probability
density $P_{opt}[\{C \}]$ is not factorized, i.e., already in the leading
order in the nonlinearity parameter, the fiber optic
channel is the channel with memory. The optimal distribution
$P_{opt}[\{C \}]$ allowed us to construct the conditional
probabilities $P_{opt}[C_k|C_{- M},\ldots, C_{k-1}, C_{k + 1},
\ldots, C_{M}]$, which, in turn, are needed to construct the input
signal with the given statistics $P_{opt}[\{C \}]$. Using the
explicit form of the distribution $P_{opt}[\{C \}]$ we demonstrated that the difference
between the mutual information found using the optimal statistics
and the mutual information calculated using the Gaussian
distribution occurs only in the fourth order in the small parameter
of the Kerr nonlinearity. To demonstrate our analytical results, we
performed the numerical calculations of mutual information, optimal
distribution function and correlators of the output signal for
various parameters of the second dispersion, as well as for pulse
sequences of different lengths.

The article is organized as follows.  The next Section is dedicated to the channel model description: we describe the structure of the input signal, then we introduce the procedures of the receiving and post-processing. We introduce the conditional probability density function in the case of small Kerr nonlinearity in the third Section. In the third Section we propose two approaches to the perturbative calculation of the conditional PDF. The details of this calculation are presented in the Appendix A. The fourth Section is devoted to the derivation of the mutual information. The resulting expression for the mutual information uses the tensor notations for the coefficients calculated in detail in the Appendix B. These universal coefficients allow us to present the optimal input signal distribution in the fifth Section. In the Section we present the theoretical and numerical results. We present the statistical method of the construction of the optimal input signal in the sixth Section: we describe the correlations of the input symbols resulting in the optimal distribution. The Conclusion finalizes our consideration of the optimal input signal distribution for the nonlinear channel with small Kerr nonlinearity.

\section{Model of the channel }

Let us start the consideration from the input signal representation.

\subsection{The input signal model}
In our model the input signal $X(t)$ has the following form
\begin{eqnarray}\label{Xtmodelg}
X(t)=\sum^{M}_{k=-M} C_{k} \, s(t-k T_0).
\end{eqnarray}
Thus, the signal is the sequence of $2M+1$ pulses of the shape
$s(t)$ spaced by time $T_0$. The complex coefficients $C_{k}$ carry
the transmitted information. We chose the pulse envelope  $s(t)$ possessing two
properties. The first property is the normalization condition:
\begin{eqnarray}
\int^{\infty}_{-\infty}\dfrac{dt}{T_0} s^2(t)=1.
\end{eqnarray}
The second property is orthogonality condition:
\begin{eqnarray}\label{nonoverlapping}
\int^{\infty}_{-\infty} \frac{dt}{T_0} s(t-k T_0) s(t-m T_0) =
\delta_{k m},
\end{eqnarray}
where $\delta_{k m}$ is Kronecker $\delta$-symbol. Below we will
consider two different types of the function $s(t)$. The first one
is the sinc-type function
\begin{equation}\label{sincft}
 s(t)=\mathrm{sinc}\left[W t/2\right]= 2\frac{\sin(W t/2)}{W t},
\end{equation}
where $W=2\pi/T_0$ is the input signal bandwidth. Note that these envelopes are overlapping, however the properties \eqref{nonoverlapping} and  \eqref{sincft} are fulfilled. We focus our attention in the following calculations primarily on the sinc type of the envelope.

The second type is the Gaussian function
\begin{equation}\label{eqInitCond}
 s(t) = \sqrt{\frac{T_0}{\tau\sqrt{\pi}}}
    \exp\left( -\frac{t^2}{2\tau^2} \right),
\end{equation}
where $\tau$ is characteristic signal duration. Below we imply that
$\tau \ll T_0$. It is the parameter $\tau$ that determines the frequency bandwidth of the input signal.  So the orthogonality condition
\eqref{nonoverlapping} can be satisfied only approximately with any specified precision by choosing the value of the time $\tau$.

The complex coefficients $C_{k}$ are distributed with probability density
function (PDF) $P_X[\{C\}]$. Below we refer the function
$P_X[\{C\}]$ as the input signal PDF, where $\{C\}=\{C_{-M},
C_{-M+1}, \ldots, C_{M}\}$ is the ordered set of the coefficients
$C_{k}$. In our model we will consider the continuous PDF
$P_X[\{C\}]$ normalized by the condition:
\begin{eqnarray}\label{normalization0}
\int \left(\prod_{k=-M}^{M} d^2 C_k\right) P_X[\{C\}]=1,
\end{eqnarray}
where $d^2 C_k= d \mathrm{Re} C_k d \mathrm{Im} C_k$. We also
restrict our consideration by the input signal $X(t)$ with the fixed
averaged power $P$:
\begin{eqnarray}\label{avpower}
P= \int \left(\prod_{k=-M}^{M} d^2 C_k\right)
P_X[\{C\}]\frac{1}{2M+1}\int^{\infty}_{-\infty}\frac{dt}{T_0} |X(t)|^2.
\end{eqnarray}

\subsection{The signal propagation model}
In our model the propagation of the signal $\psi(z,t)$  is described
by the NLSE with additive  Gaussian noise, see
{\cite{Haus:1991,Mecozzi:1994,Iannoe:1998,Turitsyn:2000}}:
\begin{eqnarray}\label{startingCannelEqt}
&&\!\!\!\partial_z \psi(z,t)+i\beta\partial^2_{t}\psi(z,t)-i\gamma
|\psi(z,t)|^2 \psi(z,t)=\eta(z,t) \,,
\end{eqnarray}
with the input condition  $\psi(0,t)=X(t)$. In
Eq.~(\ref{startingCannelEqt}) $\beta$ is the second dispersion
coefficient, $\gamma$ is the Kerr nonlinearity coefficient,
$\eta(z,t)$ is an additive complex  noise with zero mean
\begin{eqnarray}\label{zeronoisecorrelatort}
\langle \eta (z,t)\rangle_{\eta}=0.
\end{eqnarray}
Here  $\langle \ldots \rangle_{\eta}$ is the averaging
over the realization of the noise $\eta (z,t)$. We also imply that
the  correlation function $\langle \eta
(z,t)\bar{\eta}(z^\prime,t^\prime)\rangle_{\eta}$ has the following
form:
\begin{eqnarray}\label{noisecorrelatort}
\langle \eta (z,t)\bar{\eta}(z^\prime,t^\prime)\rangle_{\eta} =  Q
\frac{\tilde{W}}{2\pi}\mathrm{sinc}\left(\frac{\tilde{W}
(t-t^\prime)}{2} \right) \delta(z-z^\prime).
\end{eqnarray}
Here and below the bar means complex conjugation. The parameter $Q$
in Eq.~\eqref{noisecorrelatort} is a power of the noise $\eta(z,t)$
per unit length and per unit frequency. The parameter $\tilde{W}$ is
the bandwidth of the noise. Note that, $\lim_{\tilde{W} \rightarrow \infty} \frac{\tilde{W}}{2\pi}\mathrm{sinc}\left(\frac{\tilde{W}
(t-t^\prime)}{2} \right)=\delta(t-t')$.

Below we imply that the noise bandwidth $\tilde{W}$ is much greater
than the bandwidth $W$ of the input signal $X(t)$ and much greater
than the bandwidth $W'$ of the  solution $\Phi(z=L,t)$ of the Eq.~\eqref{startingCannelEqt} with zero noise. Here $L$ is the signal
propagation distance. So in our consideration we set that
\begin{eqnarray}\label{bandwidthscales}
\tilde{W} \gg W' > W.
\end{eqnarray}
The solution $\Phi(z,t)$ of the Eq.~\eqref{startingCannelEqt} with zero noise and with the input boundary condition $\Phi(z=0,t)=X(t)$ will play an important role in our further consideration. The details of the perturbative calculation of the solution $\Phi(z,t)$ and its properties are presented in the Subsection 2 of the Appendix A.

\subsection{The receiver model and the post-processing}

To recover the transmitted information we perform the procedure of
the output signal detection at $z=L$.  Our receiver detects the
noisy signal $\psi(L,t)$ (the solution of the
Eq.~\eqref{startingCannelEqt} with noise), then it filters the detected
signal in the frequency domain. After that we removes the phase
incursion $e^{i \beta \omega^2 L}$ related with the second
dispersion coefficient and obtain the signal ${Y}_{d}(t)$. So in the
frequency domain we finally obtain the detected signal ${Y}_{d}(\omega)$:
\begin{eqnarray} \label{Yd}
{Y}_{d}(\omega)= e^{-i \beta \omega^2 L}
\theta(W_d/2-|\omega|)\int^{\infty}_{-\infty} dt e^{i \omega t}
\psi(L,t).\,
\end{eqnarray}
where  $W_d$ is the frequency bandwidth of our receiver. In our
model the bandwidth $W_d$ is much less than $\tilde{W}$ as well as
in Eq.~\eqref{bandwidthscales}. Besides, it is reasonable to
consider the receiver with the bandwidth $W_d \geq W$, so it is our
case.

To obtain the information we should recover the coefficients $\{{C}
\}$ from the  signal ${Y}_{d}(\omega)$. To this aim we project the
signal ${Y}_{d}(t)$ on the shape functions $s(t-k T_0)$:
\begin{eqnarray}\label{tildeCk}
\tilde{C}_k&=&\frac{1}{T_0}\int^{\infty}_{-\infty} dt s(t-k
T_0){Y}_{d}(t)=\frac{1}{2 \pi T_0}  \int_{W}  {d \omega} \overline{s^{(k)}}({\omega})
{Y}_{d}({\omega}) ,
\end{eqnarray}
where $s^{(k)}(\omega)$ is the Fourier transform of the function
$s(t-k T_0)$:
\begin{eqnarray}\label{fkomega}
s^{(k)}(\omega)=\int^{\infty}_{-\infty} dt s(t-k T_0)\,e^{i \omega
t}=e^{i \omega k T_0} s({\omega}).
\end{eqnarray}
Due to the noise and nonlinearity of the
Eq.~\eqref{startingCannelEqt} the recovered coefficient
$\tilde{C}_k$ does not coincide with the coefficient $C_k$. However,
in the case of zero nonlinearity and the zero noise our detection
procedure allows us to  recover all coefficients $\{C\}$.

The informational characteristics of the channel are described by the
conditional probability density function $P[\{\tilde{C}\}|\{{C}\}]$
to receive the sequence $\{\tilde{C}\}$ for the input sequence
$\{{C}\}$. So we have to find the conditional PDF $P[\{\tilde{C}\}|\{{C}\}]$.

\section{Conditional PDF $P[\{\tilde{C}\}|\{{C}\}]$. }

In this section we find the conditional PDF
$P[\{\tilde{C}\}|\{{C}\}]$ using two approaches. The first one is
based on the result of Ref.~\cite{Terekhov:2014} where the
conditional PDF $P[Y(\omega)|X(\omega)]$ to receive the output signal $Y(\omega)$
for the input signal $X(\omega)$ was represented in the form of
path-integral. The second approach is based on the calculation of
the output signal correlators in the leading and the next-to-leading
orders in the parameter $Q$. Let us briefly discuss the first and
the second approaches.

The base of the path-integral approach is the representation  for
the conditional PDF $P[Y(\omega)|X(\omega)]$ in the frequency domain, see Ref.~\cite{Terekhov:2014}:
\begin{eqnarray}\label{pathintPYX}
P[Y(\omega)|X(\omega)]= \int^{\psi (L,\omega)=Y(\omega)}_{\psi
(0,\omega)=X(\omega)} {\cal D} \psi (z,\omega)
\exp\left[-\frac{S[\psi]}{Q}\right],
\end{eqnarray}
where the effective action $S[\psi]$ reads
\begin{eqnarray}\label{action}
S[\psi]=\int^L_0 dz \int_{\tilde{W}}
\frac{d\omega}{2\pi}\left|\partial_z \psi(z,\omega)- i\beta \omega^2
\psi(z,\omega) -i \gamma \int_{\tilde{W}} \frac{d\omega_1 d\omega_2
d\omega_3}{(2\pi)^2}\delta(\omega_1+\omega_2-\omega_3-\omega)\psi(z,\omega_1)\psi(z,\omega_2)
\bar{\psi}(z,\omega_3) \right|^2,
\end{eqnarray}
and the integration measure $ {\cal D} \psi (z,\omega)$ is defined in such a way
to obey the normalization condition $\int {\cal D} Y(\omega)
P[Y(\omega)|X(\omega)]=1$, for details see \cite{Terekhov:2014}. As
it was mentioned above, the function $P[Y(\omega)|X(\omega)]$
contains a redundant degrees of freedom, since the receiver does not
detect all frequencies of the output signal $Y(\omega)$. That is why
we have to introduce the conditional PDF
$P_{d}[Y_d(\omega)|X(\omega)]$ which is the result of the
integration of the function $P[Y(\omega)|X(\omega)]$ over  redundant
degrees of freedom $Y(\omega)$, $|\omega|>W_d/2$:
\begin{eqnarray} \label{Pdintegration}
P_{d}[Y_d(\omega)|X(\omega)]=\int [{\cal D}
Y(\omega)]_{|\omega|>W_d/2} P[Y(\omega)|X(\omega)].
\end{eqnarray}
So, the function $P_{d}[Y_d(\omega)|X(\omega)]$ contains only
detectable degrees of freedom $Y_d(\omega)$, $|\omega|<W_d/2$, see Eq.~\eqref{Yd}.
If one knows the function
$P_{d}[Y_d(\omega)|X(\omega)]$, it is easy to calculate an arbitrary
correlator $\langle \tilde{C}_{k_1}\ldots \bar{\tilde{C}}_{k_N}
\rangle$, where
\begin{eqnarray}\label{defcorrelatorsC}
\langle \tilde{C}_{k_1}\ldots \bar{\tilde{C}}_{k_N} \rangle= \int
{\cal D} Y_d(\omega) P_{d}[Y_d(\omega)|X(\omega)]
\tilde{C}_{k_1}\ldots \bar{\tilde{C}}_{k_N},
\end{eqnarray}
where $\tilde{C}_{k}$ is defined in Eq.~(\ref{tildeCk}). For our
purposes we should know correlators in the leading order in
the noise parameter $Q$, and up to the second order in the nonlinearity
parameter $\gamma$. Knowledge of these correlators allows us to
construct the conditional PDF $P[\{\tilde{C}\}|\{{C}\}]$ which
reproduces all correlators with necessary accuracy. The
details of this calculation are presented in the Appendix
\ref{AppCondPDFcalc}.

The second approach allows us to calculated the same correlators
(\ref{defcorrelatorsC}) by solving the equation
(\ref{startingCannelEqt}) up to the second order in parameter
$\gamma$ and up to the first order in the noise parameter $Q$ (i.e.,
the second order in function $\eta(z,t)$). We substitute the
solution $\psi(L,t)$ of the equation (\ref{startingCannelEqt}) to
the Eq.~(\ref{Yd}), then the result of Eq.~(\ref{Yd}) is substituted
to Eq.~(\ref{tildeCk}) and we arrive at the expression for the measured
coefficient $\tilde{C}_{k}$. Note that, since the solution
$\psi(L,t)$ depends on the noise, the coefficient $\tilde{C}_{k}$
depends on the noise function $\eta(z,t)$ as well. To calculate any
correlator $\langle \tilde{C}_{k_1}\ldots \bar{\tilde{C}}_{k_N}
\rangle$ we should average the product $\tilde{C}_{k_1}\ldots
\bar{\tilde{C}}_{k_N}$ over the  noise realizations using the Eqs.
(\ref{zeronoisecorrelatort}),  (\ref{noisecorrelatort}). As it
should be, the results for the correlators $\langle
\tilde{C}_{k_1}\ldots \bar{\tilde{C}}_{k_N} \rangle$ are the same
for both approaches. The results for the correlators are presented
in the Subsection 4 of the Appendix \ref{AppCondPDFcalc}.

Using the obtained correlators we build the conditional PDF
$P[\{\tilde{C}\}|\{{C}\}]$:
\begin{eqnarray}\label{PtildeCC0}
P[\{\tilde{C}\}|\{C\}]=\Lambda_c \exp\Big\{-\frac{T_0}{Q L}
\sum_{k,k'=-M}^{M}\Big[ \delta \tilde{C}_{k'} F^{k',k}
\overline{\delta \tilde{C}_k}+ \delta \tilde{C}_{k'} G^{k',k}
{\delta \tilde{C}_k} + \overline{\delta \tilde{C}_{k'}} H^{k',k}
\overline{\delta \tilde{C}_k}\Big] \Big\},
\end{eqnarray}
here $F^{k',k}=\bar{F}^{k,k'}=\delta^{k',k}+F^{k',k}_{2}$,
$H^{k',k}=H^{k,k'}=H^{k',k}_{1}+H^{k',k}_{2}$,
$G^{k',k}=\bar{H}^{k',k}=G^{k',k}_1+G^{k',k}_2$ are dimensionless coefficients with $k, k'=-M, \ldots ,M$. The indexes $1$ and $2$ indicate terms
proportional to $\gamma$ and $\gamma^2$, respectively. The quantity
$\delta \tilde{C}_k$ is defined as follows
\begin{eqnarray}
\delta \tilde{C}_k=\tilde{C}_k - \langle \tilde{C}_k\rangle.
\end{eqnarray}
Here the correlator $\langle \tilde{C}_k\rangle$ is known function
of $\{C\}$, see Eq.~ \eqref{avC}. Note that the  quantity $\langle \tilde{C}_k\rangle$ contains  bandwidth of the noise $\tilde{W}$.

The dimensionless coefficients can be presented through pair
correlators as
\begin{eqnarray}\label{Fcorrections0}
&&F^{k',k}=\bar{F}^{k,k'}=\delta_{k',k}+F^{k',k}_{2}, \quad
F^{k',k}_{2}=4 \sum_{m=-M}^{M} \bar{H}^{k',m}_1H^{m,k}_{1} -\gamma^2
 \frac{T_0}{2 Q L}  \frac{\partial^2}{\partial \gamma^2}\langle
\delta\bar{\tilde{C}}_{k'} \, \delta\tilde{C}_k
\rangle\Big|_{\gamma=0},
\end{eqnarray}
where
\begin{eqnarray}\label{Hkm0}
&&H^{m,k}_{1}=-\gamma\frac{T_0}{2 Q L }\frac{\partial}{\partial
\gamma}\langle \delta\tilde{C}_k \, \delta\tilde{C}_m
\rangle\Big|_{\gamma=0}, \qquad H^{m,k}_{2}=-\gamma^2\frac{T_0}{4 Q
L }\frac{\partial^2}{\partial \gamma^2}\langle \delta\tilde{C}_k \,
\delta\tilde{C}_m \rangle\Big|_{\gamma=0}.
\end{eqnarray}
All needed correlators $\langle \tilde{C}_k\rangle$, $\langle
\delta\tilde{C}_k \, \delta\tilde{C}_m \rangle$, $\langle
\delta\bar{\tilde{C}}_{k'} \, \delta\tilde{C}_k \rangle$ are
presented explicitly in the Appendix \ref{AppCondPDFcalc}: see Eqs.
\eqref{avC}, \eqref{dCdC},  \eqref{dCbardC}. The normalization
factor $\Lambda_c$ reads up to $\gamma^2$ order
\begin{eqnarray}\label{lambdac0}
\Lambda_c=\left( \frac{T_0}{\pi Q L} \right)^{2M+1} \left[1
+\left(\sum_{k=-M}^{M} F^{k,k}_{2}-2 \sum_{k,\, k\,'=-M}^{M}
G^{k,k'}_{1}H^{k',k}_{1}\right)\right].
\end{eqnarray}
At first sight, the found PDF \eqref{PtildeCC0} has the Gaussian
form, and it might be suggested that we have reduced the channel to
the linear one. But it is not the case, since the dimensionless
coefficients depends nonlinearly on the input signal coefficients
$\{C\}$. The Gaussian structure is the consequence of the
consideration of the problem in the leading order in the parameter $Q$.

Now we turn to the consideration of the channel entropies
$H[\tilde{C}]$ and $H[\{\tilde{C}\}|\{C\}]$ which are necessary for
the mutual information calculation.

\section{Mutual information}

The conditional entropy reads
\begin{eqnarray}\label{HYX}
 \!\!\! H[\{\tilde{C}\}|\{{C}\}]&=&-\int dC  d\tilde{C}  P_X[\{C\}]\, P[\{\tilde{C}\}|\{{C}\}]\, \log P[\{\tilde{C}\}|\{{C}\}],
\label{condentropy}
\end{eqnarray}
where
\begin{eqnarray}
dC=\prod^{M}_{k=-M} d \mathrm{Re}{C}_k d \mathrm{Im}{C}_k, \quad
d\tilde{C}=\prod^{M}_{k=-M} d \mathrm{Re}\tilde{C}_k d
\mathrm{Im}\tilde{C}_k.
\end{eqnarray}

To calculate the conditional entropy $H[\{\tilde{C}\}|\{C\}]$ we
substitute the conditional PDF (\ref{PtildeCC0}) to the expression
(\ref{HYX}), then perform the integration over
$\{\tilde{C}\}$, and we obtain
\begin{eqnarray}\label{HCC}
H[\{\tilde{C}\}|\{{C}\}]= -\int d{C} P_X[\{C\}] \left( \log
\Lambda_c - (2M+1)\right).
\end{eqnarray}
To perform the integration in Eq.~\eqref{HCC} we expand $\log \Lambda_c$ up to
$\gamma^2$ terms then integrate over $\{C \}$ and arrive at
\begin{eqnarray}\label{condentropCC}
H[\{\tilde{C}\}|\{C\}]=(2M+1)\log\left({\pi e Q L}/{T_0}\right)
-{\gamma^2 L^2}  J^{s_1,s_2;{s_3},{s_4}}_{\Lambda} \int  d{C}
P_{X}[\{C\}] {C}_{s_1}{C}_{s_2}\bar{{C}}_{s_3}\,\,\bar{{C}}_{s_4}.
\end{eqnarray}
To obtain Eq.~\eqref{condentropCC} we have used the normalization condition
\begin{eqnarray}
\int d{C}  P_X[\{C\}]=1.
\end{eqnarray}
In Eq.~\eqref{condentropCC} and below, unless otherwise stated, we imply that there is the
summation over the repeated indices. The explicit expression for  coefficients
$J^{s_1,s_2;{s_3},{s_4}}_{\Lambda}$ is cumbersome, therefore we
present it in the Appendix \ref{AppCondPDFcalc}, see
Eq.~\eqref{Jlambda1}.

Now we proceed to the calculation the output signal entropy
\begin{eqnarray}\label{HY}
\!\!\! H[\{\tilde{C}\}]&=&-\int  d\tilde{C} P_{out}[\{\tilde{C}\}]
\log P_{out}[\{\tilde{C}\}], \label{entropies}
\end{eqnarray}
where the output signal distribution reads
\begin{eqnarray}\label{Pout}
P_{out}[\{\tilde{C}\}]= \int   d{C} P[\{\tilde{C}\}|\{{C}\}]
P_X[\{C\}].
\end{eqnarray}
To calculate the PDF of the output signal
 we change the integration variables in
Eq.~\eqref{Pout} from $C_k$ to $\delta \tilde{C}_k=\tilde{C}_k -
\langle \tilde{C}_k\rangle$.

Since in our model the average noise power is much less than the
average input signal power ($QL/T_0 \ll P$), we calculate the
integral (\ref{Pout}) using the Laplace method
\cite{Lavrentiev:1987} and obtain the following result in the
leading order in the  parameter $1/\mathrm{SNR}=QL/(T_0 P)$:
\begin{eqnarray}\label{PoutCresult}
P_{out}[\{\tilde{C}\}] \approx \left|\frac{\partial
(C,\bar{C})}{\partial
(\tilde{C}^{(0)},\bar{\tilde{C}}^{(0)})}\right| P_X[\{F\}],
\end{eqnarray}
where $\tilde{C}^{(0)}_k$ is the known function of ${C}_k$, see  Eq.
\eqref{ctildezero} in Appendix A.
\begin{eqnarray}\label{ctildezero0}
&& \tilde{C}^{(0)}_k[\{C\}]={C}_k+ i{\gamma L}
{C}_{k_1}{C}_{k_2}\bar{{C}}_{k_3}a^{k_1,k_2;{k_3},{k}}_{1} -
{\gamma^2 L^2}
{C}_{m_1}{C}_{m_2}{C}_{m_3}\bar{{C}}_{m_4}\,\bar{{C}}_{m_5}a^{m_1,m_2,m_3;\,{m_4},{m_5},{k}}_2,
\end{eqnarray}
and  $F[\{\tilde{C}\}]$ is the  solution of the equation
\begin{eqnarray}\label{F}
&& \tilde{C}_k=\tilde{C}^{(0)}_k[\{F\}].
\end{eqnarray}
The solution $F$ of this equation can be found using perturbation
theory in the parameter $\gamma$. One can see, that the distribution
of the output signal coincides  with the input signal distribution
$P_X[\{F\}]$ up to the Jacobian determinant $\left|\frac{\partial
(C,\bar{C})}{\partial
(\tilde{C}^{(0)},\bar{\tilde{C}}^{(0)})}\right|$. It allows us to
calculate the output signal entropy in the leading order in
parameter $Q$, or $1/\mathrm{SNR}=QL/(T_0 P)$ in the dimensionless
quantities.

Substituting the result (\ref{PoutCresult}) into the expression for
the output signal entropy \eqref{HY}, we perform the integration
over $\tilde{C}$ and arrive at
\begin{eqnarray}\label{HtildeC}
H[\{\tilde C\}]=H[\{ C\}]+ \int d C  P_X[\{C\}] \log
\left|\frac{\partial
(\tilde{C}^{(0)},\bar{\tilde{C}}^{(0)})}{\partial
(C,\bar{C})}\right|,
\end{eqnarray}
where $H[\{ C\}]$ is the input signal entropy:
\begin{eqnarray}\label{HC}
H[\{ C\}]=- \int d C  P_X[\{C\}] \log P_X[\{C\}].
\end{eqnarray}
Therefore to find the output entropy we should calculate the
logarithm of the Jacobian determinant in Eq.~(\ref{HtildeC}). The
straightforward calculation in the first non-vanishing  order in the
parameter $\gamma$ leads to the following result:
\begin{eqnarray}\label{Jacobian}
&& \log \left|\frac{\partial
(\tilde{C}^{(0)},\bar{\tilde{C}}^{(0)})} {\partial
(C,\bar{C})}\right|= {\gamma^2 L^2}
{C}_{s_1}{C}_{s_2}\bar{{C}}_{s_3}\,\,\bar{{C}}_{s_4}  J^{s_1,s_2;
{s_3},{s_4}},
\end{eqnarray}
where dimensionless coefficients $J^{s_1,s_2;{s_3},{s_4}}$ are given
by the Eq.~\eqref{JJacobian4}.

To calculate the mutual information we subtract the conditional
entropy (\ref{condentropCC}) from the output signal entropy
(\ref{HtildeC}):
\begin{eqnarray}\label{MI}
&&
I_{P_X}=H[\{\tilde{C}\}]-H[\{\tilde{C}\}|\{C\}]=-(2M+1)\log\left[{\pi
e Q L}/{T_0}\right]-\int d C P_{X}[\{C\}]\log P_{X}[\{C\}]
+\nonumber \\&& {\gamma^2 L^2} J_{I}^{s_1,s_2;{s_3},{s_4}} \int d C
P_{X}[\{C\}] {C}_{s_1}{C}_{s_2}  \bar{{C}}_{s_3}\,\,\bar{{C}}_{s_4},
\end{eqnarray}
where coefficients
\begin{eqnarray}
J_{I}^{s_1,s_2;{s_3},{s_4}}=J^{s_1,s_2;{s_3},{s_4}}+J_{\Lambda}^{s_1,s_2;{s_3},{s_4}}.
\end{eqnarray}
These two contributions to the coefficients
$J_{I}^{s_1,s_2;{s_3},{s_4}}$ are presented explicitly in
Eqs.~(\ref{JJacobian4}) and (\ref{Jlambda}) . The method of the
numerical calculation of these coefficients is presented in Appendix
\ref{NumCalcSection}. The first two terms in the mutual information
(\ref{MI}) coincide with that for linear channel ($\gamma=0$). The
third term describes the contribution of the Kerr nonlinearity
effects. One can see that the first nonlinear correction to the
mutual information is of the order of $\gamma^2$. Since the
coefficients $J_{I}^{s_1,s_2;{s_3},{s_4}}$ depend on the envelope
function $s(t)$, therefore the mutual information also depends on
this envelope function. It is worth noting, that the mutual
information depends on the bandwidth of the input signal $W$ via
coefficients $J_{I}^{s_1,s_2;{s_3},{s_4}}$  and does not depend on
the detector bandwidth $W_d$. The reason for that is the bandwidth
$W_d$ of the receiver is greater than or equal to the bandwidth of
the input signal $W$: $W_d \geq W$. Therefore, all integrals over
frequency in the interval $[-W_d/2, W_d/2]$ with the envelope
$s^{(k)}(\omega)$ are reduced to the integrals over the frequency
interval $[-W/2, W/2]$ determined by the function $s({\omega})$.

\section{Optimal input signal distribution}

Now we can calculate the optimal input signal distribution
$P_{opt}[\{C\}]$ which maximizes the mutual information (\ref{MI}).
The optimal distribution  $P_{opt}[\{C\}]$ obeys the normalization
condition:
\begin{eqnarray}\label{normalization1}
\int d C P_{opt}[\{C\}]=1,
\end{eqnarray}
and the condition of the fixed average power:
\begin{eqnarray}\label{normalization2}
\int d C P_{opt}[\{C\}] \frac{1}{2M+1}\sum_{k=-M}^M|C_k|^2=P.
\end{eqnarray}

To find $P_{opt}[\{C\}]$ we solve the variational problem $\delta
\,{\cal K}[P_X] = 0$ for the functional
\begin{eqnarray}
{\cal K}[P_X]=I_{P_X}-\lambda_1 \left(\int {\cal D}C
P_X[\{C\}]-1\right)-\lambda_2 \left(\int {\cal D}C P_X[\{C\}]
\frac{1}{2M+1}\sum_{k=-M}^M|C_k|^2-P\right),
\end{eqnarray}
where $\lambda_{1,2}$ are the Lagrange multipliers, related with the
restrictions \eqref{normalization1}, \eqref{normalization2}. The
solution of the variational problem in the first and in the second
orders in the parameter $\gamma$ and in the leading order in the
parameter $1/\mathrm{SNR}$ has the form:
\begin{eqnarray}\label{Popt}
&&P_{opt}[\{C\}]=P^{(0)}[\{C\}]\Big\{1+\gamma^2 L^2
J_{I}^{s_1,s_2;{s_3},{s_4}}{C}_{s_1}{C}_{s_2}\bar{{C}}_{s_3}\,\,\bar{{C}}_{s_4}+
\nonumber \\&& \left(\gamma L P\right)^2
\left(J_{I}^{r,s;{r},{s}}+J_{I}^{r,s;{s},{r}}\right)
\Big(1-\frac{2}{P(2M+1)}\sum_{k=-M}^{M}|C_k|^2\Big)\Big\},
\end{eqnarray}
where $P^{(0)}[\{C\}]$ is the optimal input signal distribution for
the channel with zero nonlinearity parameter $\gamma$:
\begin{eqnarray}\label{Poptzerogamma}
P^{(0)}[\{C\}]=\left(\frac{1}{\pi
P}\right)^{2M+1}\exp\left[-\frac{1}{P}\sum^M_{k=-M}|C_k|^2\right].
\end{eqnarray}
One can see, that $P^{(0)}[\{C\}]$ is the Gaussian distribution,
whereas the distribution (\ref{Popt}) is not Gaussian due to  the
nonlinear corrections. Thus, $P_{opt}[\{C\}]$ leads to the nonzero
correlations between coefficients $C_k$ with different $k$.

Note that the found distribution $P_{opt}[\{C\}]$ is not the exact
optimal input signal distribution for the channel which is described
by the nonlinear Schr\"odinger  equation, since $P_{opt}[\{C\}]$ is
calculated for the given envelope $s(t)$ and only up to the
$\gamma^2$ terms. Nevertheless, $P_{opt}[\{C\}]$ takes into account
the first nonzero nonlinear corrections that lead to nontrivial
correlations of the input coefficients $C_k$.

To find the maximal value of the mutual information (\ref{MI}) we
substitute  $P_{opt}[\{C\}]$ (\ref{Popt}) to the expression
(\ref{MI}), perform the integration over $C$ and obtain
\begin{eqnarray}\label{IPopt}
I_{P_{opt}}=(2M+1)\left(\log \left[\frac{PT_0}{Q
L}\right]+\left(\gamma L P\right)^2
J_{\Sigma}\right),
\end{eqnarray}
where
\begin{eqnarray}\label{Jsigma}
J_{\Sigma}=\frac{{J_{I}^{r,s;{r},{s}}+J_{I}^{r,s;{s},{r}}}}{2M+1}.
\end{eqnarray}
One can see that the  mutual information is proportional to the
number of the coefficients $C_k$, i.e. $2M+1$. The first term in the
second brackets  in  Eq.~(\ref{IPopt}) coincides with the Shannon's
result for the linear channel at large $\mathrm{SNR}$. The second
term is the first nonzero nonlinear correction. Below we will
demonstrate numerically  that the quantity $J_{\Sigma}$ depends weakly on $2M+1$.

We emphasize that the calculation of the mutual information
(\ref{MI}) using the Gaussian distribution (\ref{Poptzerogamma})
leads to the result $I_{P^{(0)}}$ which coincides with the result
(\ref{IPopt}). It means that in this order in the parameter $\gamma$
both distributions give the same result for the mutual information.
Thus, one might think that it doesn't matter what distribution, the
optimal \eqref{Popt} or the Gaussian \eqref{Poptzerogamma}, is used
for the calculation of the mutual information. However, the optimal
input signal distribution (\ref{Popt}) leads to the mutual
information $I_{P_{opt}}$ that is greater than $I_{P^{(0)}}$ in the
higher orders in the nonlinearity parameter $\gamma$. To demonstrate
that we have calculated the difference between $I_{P_{opt}}$ and
$I_{P^{(0)}}$ in the leading nonzero order in the nonlinearity
parameter $\gamma$ and obtain:
\begin{eqnarray}\label{deltaI}
I_{P_{opt}}-I_{P^{(0)}}= \frac{(\gamma L P)^4}{2} \left(\langle A^2
\rangle_{P^{(0)}}- \langle A
\rangle^2_{P^{(0)}}-\frac{4}{2M+1}\langle A \rangle^2_{P^{(0)}}
\right), \end{eqnarray} where
\begin{eqnarray}
A=\frac{J_{I}^{s_1,s_2;{s_3},{s_4}} }{P^2}{C}_{s_1}{C}_{s_2}
\bar{{C}}_{s_3}\,\,\bar{{C}}_{s_4},
\end{eqnarray}
and here we introduce the averaging over the zero-order distribution
\eqref{Poptzerogamma}:
\begin{eqnarray}
\langle (\ldots) \rangle_{P^{(0)}}= \int d C P^{(0)}[\{C\}](\ldots).
\end{eqnarray}
Performing the averaging in Eq.~(\ref{deltaI}) we arrive at the
following  result
\begin{eqnarray}\label{deltaIexpl}
I_{P_{opt}}-I_{P^{(0)}}= 2{(\gamma L P)^4}\left(4
\tilde{J}_{I}^{a,b;b,d}\tilde{J}_{I}^{d,k;a,k}+
\tilde{J}_{I}^{a,b;c,d} \tilde{J}_{I}^{c,d;a,b}-\frac{4}{2M+1}
\tilde{J}_{I}^{a,b;b,a} \tilde{J}_{I}^{c,d;d,c}  \right),
\end{eqnarray}
where
\begin{eqnarray}
\tilde{J}_{I}^{a,b;c,d}=\left({J}_{I}^{a,b;c,d}+{J}_{I}^{b,a;c,d}+{J}_{I}^{a,b;d,c}+{J}_{I}^{b,a;d,c}\right)/4.
\end{eqnarray}
We have checked numerically that the right-hand side of
\eqref{deltaIexpl} is positive for all considered in the present
paper dispersions $\beta$.

Below we present results for the mutual information for different
envelopes $s(t)$ and for different values of dispersion.

\subsection{Zero dispersion case}

The direct calculation of the  mutual information \eqref{IPopt} in
the case of zero dispersion and non-overlapping envelopes $s(t)$,
obeying the condition \eqref{nonoverlapping}, see for instance  the
envelope~\eqref{eqInitCond}, gives the result
\begin{eqnarray}\label{IPzero}
\left.\frac{I_{P_{opt}}}{2M+1}\right|_{\beta=0}=\log
\left[\frac{PT_0}{Q L}\right]-\left(\gamma L
P\right)^2\frac{22N_6-21N^2_4}{3},
\end{eqnarray}
where $N_\lambda$ is the integral
\begin{eqnarray}
N_\lambda=\frac{1}{T_0}\int^{\infty}_{-\infty} dt s^\lambda(t).
\end{eqnarray}
We note that $22N_6-21N^2_4 > 0$ due to Cauchy-Schwarz-Bunyakovsky
inequality. For the case of the rectangular pulse
$s(t)=\theta(T_0/2-|t|)$  (which corresponds to the case of the
per-sample model, see \cite{Terekhov:2016a}) the Eq.~(\ref{IPzero})
passes to
\begin{eqnarray}\label{IPzerorec}
\left.\frac{I_{P_{opt}}}{2M+1}\right|_{\beta=0}=\log
\left[\frac{PT_0}{Q L}\right]-\frac{\left(\gamma L P\right)^2}{3}.
\end{eqnarray}
This result coincides with Eq.~(53) in Ref.~\cite{Terekhov:2016a}.

The difference \eqref{deltaI} for the case $\beta=0$ and
non-overlapping envelopes $s(t)$, see Eq.~\eqref{eqInitCond},  has
the form
\begin{eqnarray}
\left.I_{P_{opt}}-I_{P^{(0)}}\right|_{\beta=0}=(2M+1)(\gamma L P)^4
\frac{(22N_6-21N^2_4)^2}{18}.
\end{eqnarray}
One can see that the difference is positive and in agreement with
the general results of Ref.~\cite{Terekhov:2016a} that is valid for
the arbitrary nonlinearity.

For the case of the envelope of the sinc form,
\begin{equation}\label{Sincft}
 s(t)=\mathrm{sinc}\left(W t/2\right),
\end{equation}
we obtain the following result for the maximum value of the mutual
information \eqref{IPopt}, see details in the Appendix B, subsection 3:
\begin{eqnarray}\label{IPoptbetazero}
\frac{I_{P_{opt}}}{(2M+1)}&=&\log \left[\frac{PT_0}{Q
L}\right]+\frac{\left(\gamma L P\right)^2}{2M+1}
\Big(-\frac{22}{3}\int^{+\infty}_{-\infty} d \tau
S^6(\tau,\tau)+\nonumber \\&& \int^{+\infty}_{-\infty} d \tau_1
\int^{+\infty}_{-\infty} d \tau_2 \left(3
S^8(\tau_1,\tau_2)+4S^4(\tau_1,\tau_2)S^2(\tau_1,\tau_1)S^2(\tau_2,\tau_2)
\right) \Big),
\end{eqnarray}
where
\begin{eqnarray}
S^2(\tau_1,\tau_2)=\sum^{M}_{r=-M}\mathrm{sinc}(\pi(\tau_1+r))\mathrm{sinc}(\pi(\tau_2+r)),
\qquad S^2(\tau,\tau)=\sum^{M}_{r=-M}\mathrm{sinc}^2(\pi(\tau+r)).
\end{eqnarray}
The numerical result for the quantity \eqref{IPoptbetazero} has the
form
\begin{eqnarray}\label{IPoptbetazeronum}
\frac{I_{P_{opt}}}{(2M+1)}&=&\log \left[\frac{PT_0}{Q L}\right]-1.26
\left(\gamma L P\right)^2,
\end{eqnarray}
where the coefficient at the nonlinearity factor $\left(\gamma L
P\right)^2$ weakly depends on the parameter $M$.

\subsection{Nonzero dispersion case}

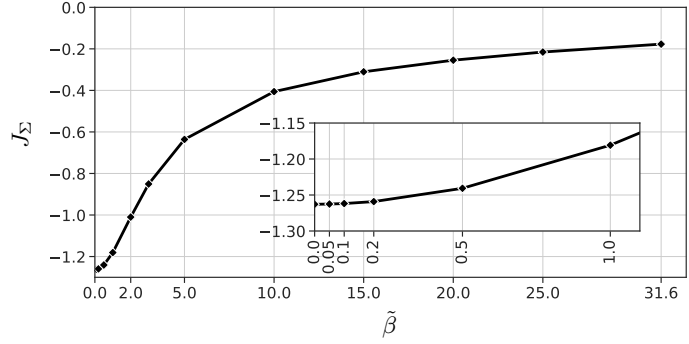
\begin{figure}[t]
    \centering
    \includegraphics[width=0.50\linewidth]{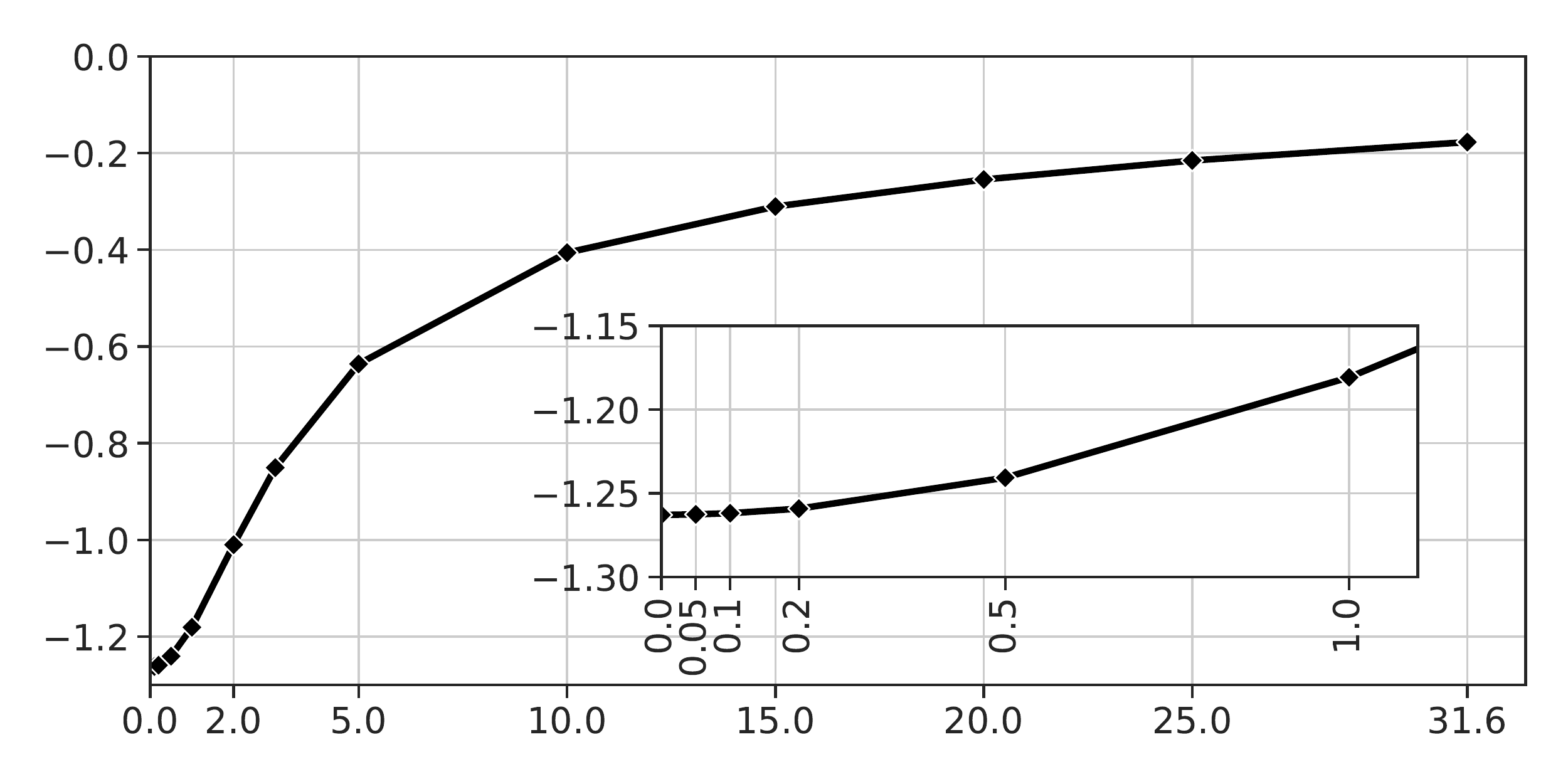}
    \begin{picture}(0,0)
\put(-125,-5){$\tilde{\beta}$} \put(-265,65){\rotatebox{90}{$J_{\Sigma}$}}
\end{picture}
    \caption{The dependence of
$J_{\Sigma}$ on the dispersion parameter
$\tilde{\beta}$ for $M=5$. In the inset we plot
$J_{\Sigma}$ for the region of small $\tilde{\beta}$.}
    \label{figPoptI}
\end{figure}

Here we present the results for the nonzero dispersion parameter.
For this consideration we choose the following  parameters of the
channel: $\beta= 2\times 10^{-23} \mathrm{sec}^2/\mathrm{km} $, the
propagation length is equal to $L=800\, \mathrm{km}$, and different
values of the input signal bandwidth $W$. The dispersion effects can
be described by the dimensionless parameter  $\tilde{\beta}$:
\begin{eqnarray}\label{tildebeta}
\tilde{\beta}=\beta L W^2/2.
\end{eqnarray}
Below we present the numerical results for the mutual information
for various values of $\tilde{\beta}$.  Fig.~\ref{figPoptI} presents
the dependence of the quantity
$J_{\Sigma}$, see Eq.~\eqref{Jsigma}, on
different values of the parameter $\tilde{\beta}$ for $M=5$, see
Eq.~\eqref{IPopt}. We checked that the quantity $J_{\Sigma}$ weakly
depends on $M$ for $M>5$.  The points in the Fig.~\ref{figPoptI}
were obtained by two different numerical approaches, see Appendix B.
Both approaches leads to the same results and it is the guarantee of
correctness of the numerical calculations.
One can see that $J_{\Sigma}$ has the minimum at $\tilde{\beta}=0$.
It means that nonlinear correction to the mutual information, see
\eqref{IPopt}, has the maximum absolute value at $\tilde{\beta}=0$.
At small  $\tilde{\beta}$ the quantity  $J_{\Sigma}$ demonstrates
the quadratic dependence on the dispersion parameter, see the inset
in Fig.~\ref{figPoptI}. At large $\tilde{\beta}$ the quantity
$J_{\Sigma}$ goes to zero. Our estimations results in the dependence
$J_{\Sigma} \sim \frac{\log \tilde{\beta}}{\tilde{\beta}}$ for large
$\tilde{\beta}$, see \cite{Terekhov:2016b}. Therefore, the nonlinear
correction decreases with increasing $\tilde{\beta}$, as ${(\gamma L
P)^2}/{\tilde{\beta}}$. It means that effective nonlinear parameter
at large  $\tilde{\beta}$ is not $\gamma L P$, but $\gamma L
P/\sqrt{{\tilde{\beta}}}$.

\section{Construction of the input signals}
\label{SectionConstruction} To increase the mutual information we
should be able to create the signals which obey the optimal input
signal distribution \eqref{Popt}. To create the input sequence which
has the statistics determined by the PDF $P_{opt}[\{C\}]$, we
represent this PDF in  the form \cite{Voitishek}:
\begin{eqnarray}
P_{opt}[\{C\}]=P_{opt}[C_{i_1}]\times P_{opt}[C_{i_2}|C_{i_1}]\times
P_{opt}[C_{i_3}|C_{i_2},C_{i_1}]\times\ldots \times
P_{opt}[C_{i_{2M+1}}|C_{i_{2M}},\ldots C_{i_2},C_{i_1}],
\end{eqnarray}
where
\begin{eqnarray}
P_{opt}[C_{i_1}]&=&\int{d}C_{i_2}\ldots{d}C_{i_{2M+1}}P_{opt}[\{C\}],\label{PoptC1}\\
P_{opt}[C_{i_2}|C_{i_1}]&=&\frac{\int{d}C_{i_3}\ldots{d}C_{i_{2M+1}}P_{opt}[\{C\}]}{P_{opt}[C_{i_1}]},\label{P[CC]General}\\
P_{opt}[C_{i_3}|C_{i_2},C_{i_1}]&=&\frac{\int{d}C_{i_4}\ldots{d}C_{i_{2M+1}}P_{opt}[\{C\}]}{P_{opt}[C_{i_1}]\times P_{opt}[C_{i_2}|C_{i_1}]},\\
P_{opt}[C_{i_{2M}}|C_{i_{2M-1}},\ldots C_{i_2},C_{i_1}]&=&\frac{\int{d}C_{i_{2M+1}}P_{opt}[\{C\}]}{P_{opt}[C_{i_1}]\times\ldots\times P_{opt}[C_{i_{2M-1}}|C_{i_{2M-2}},\ldots,C_{i_1}]},\\
P_{opt}[C_{i_{2M+1}}|C_{i_{2M}},\ldots
C_{i_2},C_{i_1}]&=&\frac{P_{opt}[\{C\}]}{P_{opt}[C_{i_1}]\times\ldots\times
P_{opt}[C_{i_{2M}}|C_{i_{2M-1}},\ldots,C_{i_1}]}.\label{PoptCM}
\end{eqnarray}
Using Eqs. \eqref{PoptC1}-\eqref{PoptCM} we can build the sequences
which have the necessary statistics by the following way. At first,
we choose the first element $C_1$ of the sequence distributed with
PDF \eqref{PoptC1}. The statistics of the second element $C_2$
depends on the value of $C_1$, and should be distributed with PDF
\eqref{P[CC]General}, etc. In our approximation ($\gamma^2$ order of
the calculation) the optimal PDF $P_{opt}[\{C\}]$ contains  the
fourth order polynomial in the coefficients $C_k$. So, we have two
nontrivial correlators $\langle C_k\bar{C}_m\rangle_{P_{opt}}$, and
$\langle C_k C_q \bar{C}_m \bar{C}_p\rangle_{P_{opt}}$, which
determine all higher order correlators.

For a very long sequence the correlation between the first and the
last coefficients should be neglectable. To find the characteristic
length $|k-m|$ of the correlation between elements $C_k$ and $C_m$
of the input sequence, we calculate the correlator $\langle
C_k\bar{C}_m\rangle_{P_{opt}}$. After the straightforward
calculation we obtain:
\begin{eqnarray}
&&\langle C_k\bar{C}_m\rangle_{P_{opt}}=P\delta_{km}\left(1-(\gamma
L P)^2\frac{2}{2M+1}\sum^M_{r,s=-M}
\left(J_{I}^{r,s;{r},{s}}+J_{I}^{r,s;{s},{r}}\right)\right)+\nonumber\\&&
P(\gamma L P)^2
\sum_{r=-M}^M\left[J_{I}^{r,m;{r},{k}}+J_{I}^{r,m;{k},{r}}+J_{I}^{m,r;{r},{k}}+J_{I}^{m,r;{k},{r}}\right].
\end{eqnarray}
The first term in the right-hand side of this equation contains the
Kronecker delta-symbol, i.e. it is zero for $k \neq m$.  The second
term describes the correlation between different elements of the
input sequence. To find the correlation  length we should
investigate the dependence of this term on the parameter $m-k$. The
correlation  length depends on the parameter  $\tilde{\beta}$. For
the small $\tilde{\beta}$ only the nearest neighbors are correlated
since the spreading of the input signal due to the dispersion is
small. When increasing the dispersion parameter $\tilde{\beta}$, the
correlation  length is increasing. The numerical values of the
coefficients $J_{I}^{i,j;{k},{l}}$ are presented in Supplementary
materials. So, one can calculate any necessary correlators.


As an example, we consider the sequence where only the nearest
elements are correlated. To build the sequence we should know only
two distributions: $P_{opt}[C_i]$ and
$P_{opt}[C_i|C_j]=P_{opt}[C_i,C_j]/P_{opt}[C_j]$. We have performed
the calculation of these distributions and obtain:
\begin{eqnarray}\label{PoptCq}
&&P_{opt}[C_q]=P^{(0)}[C_q]\left(1+(\gamma L P)^2
D^{(q)}_1\left({|C_q|}/{\sqrt{P}}\right)\right),
\end{eqnarray}
where $P^{(0)}[C_q]=\frac{1}{\pi P}
\exp\left\{-\frac{|C_q|^2}{P}\right\}$ is the Gaussian distribution,
and $D^{(q)}_1(x)$ is the following polynomial function:
\begin{eqnarray}\label{PoptDx}
D^{(q)}_1(x)&=&
\Bigg[(1-x^2)\left(2\sum^M_{r,s=-M}\frac{J_{I}^{r,s;{r},{s}}+J_{I}^{r,s;{s},{r}}}{2M+1}-\sum_{r=-M}^M
\left[J_{I}^{r,q;{r},{q}}+J_{I}^{r,q;{q},{r}}+J_{I}^{q,r;{r},{q}}+J_{I}^{q,r;{q},{r}}\right]\right)+\nonumber\\&&
J_I^{q,q;q,q}\left(x^4-4x^2+2\right)\Bigg],
\end{eqnarray}

\begin{eqnarray}\label{P[CC2]}
&&P_{opt}[C_i,C_j]=P_{opt}[C_j,C_i]=\int{d}C_{i_3}\ldots{d}C_{i_{2M+1}}P_{opt}[\{C\}]=\nonumber
\\&& P_{opt}[C_i]P_{opt}[C_j]\left\{ 1+(\gamma P L)^2
D^{i,j}\left(\frac{C_i}{\sqrt{P}},\frac{C_j}{\sqrt{P}}\right)
\right\},
\end{eqnarray}
\begin{eqnarray}
&& D^{i,j}(x,y)= J_I^{i,i;j,j}{x^2\bar{y}^2}+J_I^{j,j;i,i}{\bar{x}^2y^2}+\left({|x|^2}-1\right) \left({|y|^2}-1\right) \left(J_I^{i,j;i,j}+J_I^{i,j;j,i}+J_I^{j,i;i,j}+J_I^{j,i;j,i}\right)+ \nonumber\\
&& {x \bar{y}}\Big\{ \left(J_I^{i,i;i,j}+J_I^{i,i;j,i}\right)\left({|x|^2}-2\right)+ \left(J_I^{i,j;j,j}+J_I^{j,i;j,j}\right)\left({|y|^2}-2\right) \Big\}+ \nonumber\\
&& {y\bar{x}}\Big\{ \left(J_I^{i,j;i,i}+J_I^{j,i;i,i}\right)\left({|x|^2}-2\right)+  \left(J_I^{j,j;i,j}+J_I^{j,j;j,i}\right)\left({|y|^2}-2\right) \Big\}+\nonumber\\
&&\sum_{m=-M}^M\left({x\bar{y}}\left[J_I^{i,m;j,m}+J_I^{i,m;m,j}+J_I^{m,i;j,m}+J_I^{m,i;m,j}\right]+{\bar{x}y}\left[J_I^{j,m;i,m}+J_I^{j,m;m,i}+J_I^{m,j;i,m}+J_I^{m,j;m,i}\right]
\right).
\end{eqnarray}

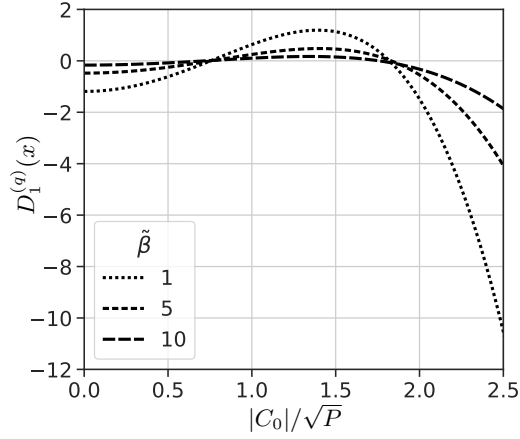
\begin{figure}[h]
    \centering
    \includegraphics[width=0.4\linewidth]{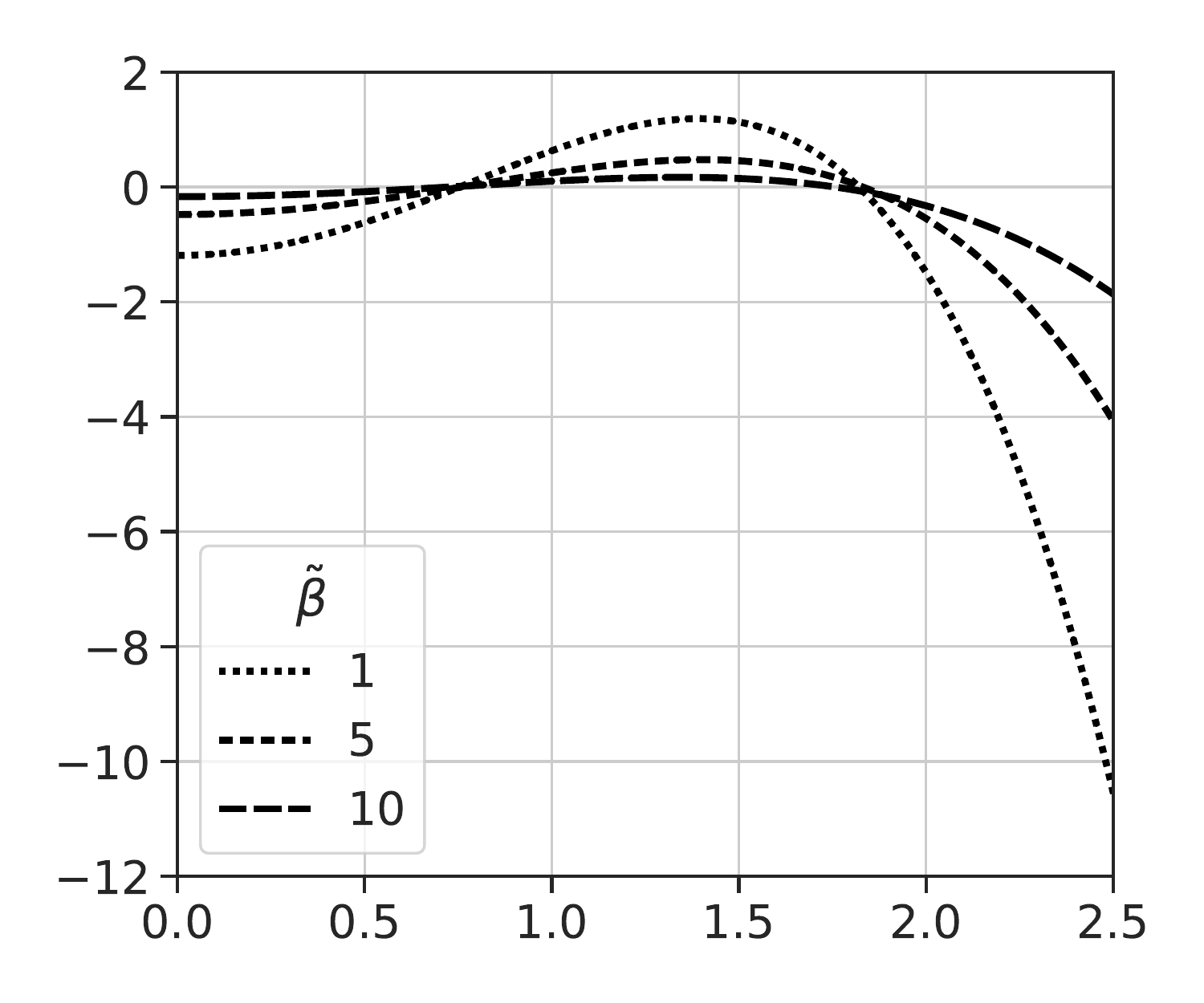}
    \begin{picture}(0,0)
\put(-115,0){$|C_0|/\sqrt{P}$} \put(-205,80){\rotatebox{90}{$D^{(q)}_1(x)$}}
\end{picture}
    \caption{The dependence of the function $D^{(q)}_1(x)$ on $x=|C_0|/\sqrt{P}$ for $q=0$ and  different values of the dispersion parameter $\tilde{\beta}$. The dotted, dashed,
long-dashed lines are plotted for the dispersion parameter
$\tilde{\beta}$ equals to $\tilde{\beta}=1$,  $\tilde{\beta}=5$,
$\tilde{\beta}=10$, correspondingly.}
    \label{figPopt1}
\end{figure}

One can see that the corrections to these PDFs are the fourth order
polynomials in parameter  $C_q/\sqrt{P}$. Let us consider these
polynomials.  In Fig. \ref{figPopt1} we plot the function $D^{(q)}_1(x)$,
see Eq. \eqref{PoptDx}, for  different values of $\tilde{\beta}$.
The function $D^{(q)}_1(x)$ for different  $\tilde{\beta}$ has the maximum
in the vicinity of the value $x \approx 1.5$. For  $x > 1.5$ this
function decreases for all values of  $\tilde{\beta}$. For smaller
$\tilde{\beta}$ the absolute value of the function  $D^{(q)}_1(x)$ is
larger for  $x > 2$. It means the applicability region determined
by the relation $(\gamma L P)^2 D^{(q)}_1({|C_q|}/{P}) \ll 1$  is wider for larger
$\tilde{\beta}$. The reason is the decreasing character of
the coefficients $J_I^{i,j;k,l}$ for increasing $\tilde{\beta}$, see
e.g. Fig.~\ref{figPopt1}.

\begin{figure}[h]
    \centering
    \includegraphics[width=0.4\linewidth]{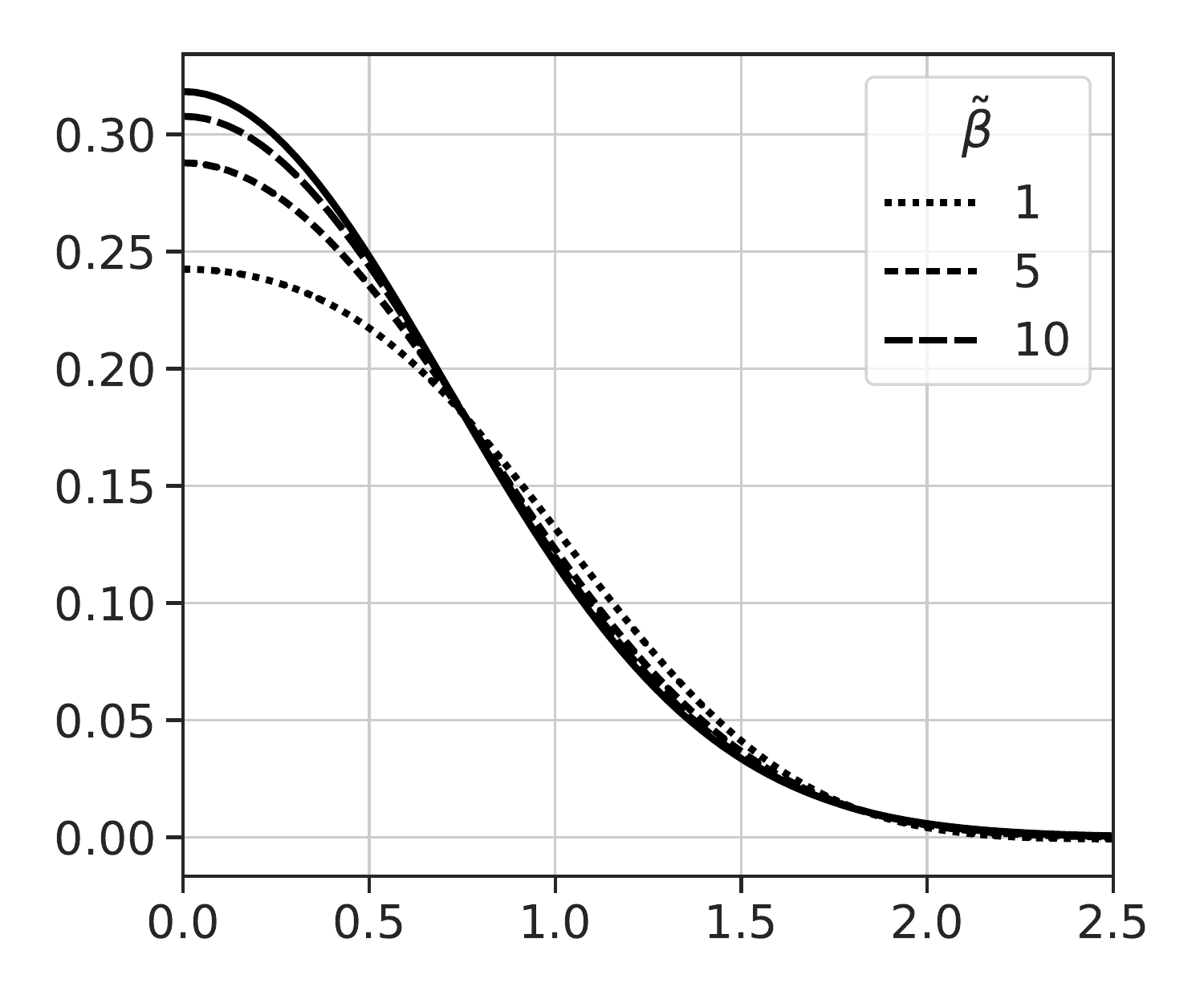} 
    \begin{picture}(0,0)
\put(-115,0){$|C_0|/\sqrt{P}$} \put(-210,80){\rotatebox{90}{$P_{opt}[C_0]$}}
\end{picture}
\caption{The function  $P_{opt}[C_0]$ for different values of the
dispersion $\tilde{\beta}$. The solid line corresponds to the
Gaussian distribution $P^{(0)}[C_0]$ for the power parameter $P=1$
in conventional units. The dotted, dashed,
long-dashed lines are plotted for the dispersion parameter
$\tilde{\beta}$ equals to $\tilde{\beta}=1$,  $\tilde{\beta}=5$,
$\tilde{\beta}=10$, correspondingly.
}
    \label{figPopt2}
\end{figure}
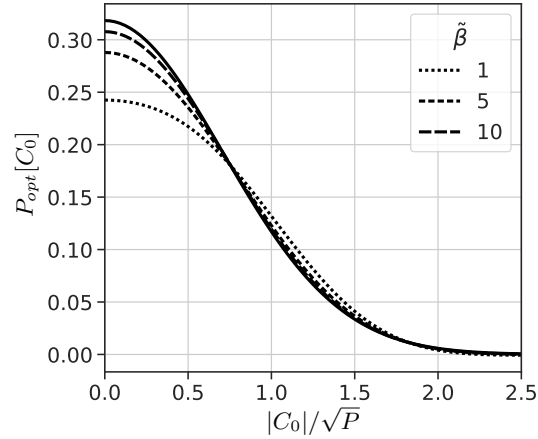

To demonstrate the behavior of the function  $P_{opt}[C_q]$ we plot
it for the different dispersion parameter  $\tilde{\beta}$, see Fig.~\ref{figPopt2}.
In Fig.~\ref{figPopt2} we chose
the nonlinear parameter $(\gamma L P)^2=0.2$.
One can see that the function $P_{opt}[C_0]$ decreases slowly for
smaller $\tilde{\beta}$. It means that nonlinear correction
decreases with increasing $\tilde{\beta}$. The difference
$|P^{(0)}[C_0]-P_{opt}[C_0]|$ is getting smaller when increasing
$\tilde{\beta}$. The reason is that for the larger dispersion
parameter the signal spreading is larger. It results in the decreasing
of the effective nonlinearity parameter, i.e., decreasing the
coefficients $J_I^{i,j;k,l}$.

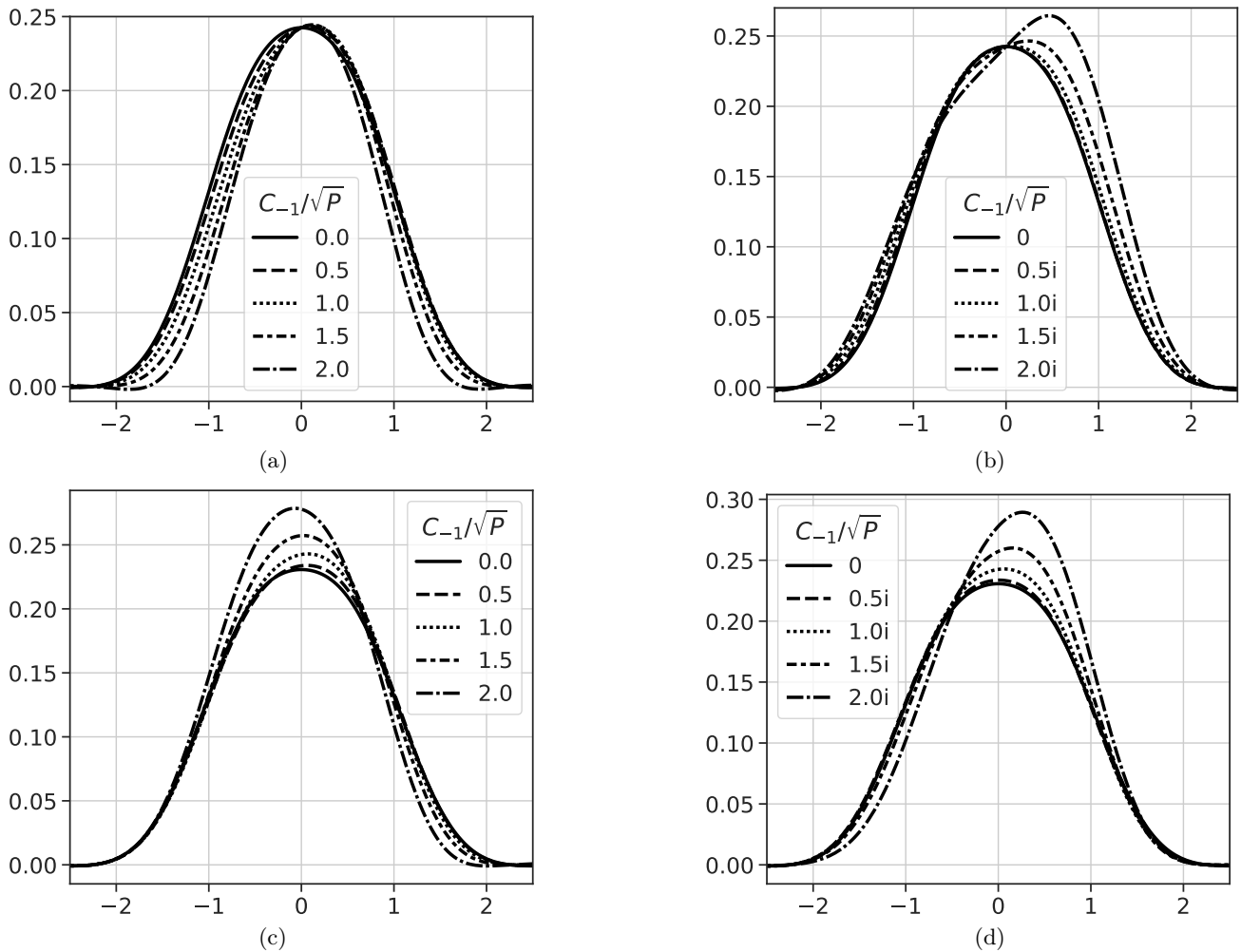
\begin{figure}[t]
    \centering
    \includegraphics[width=0.45\linewidth]{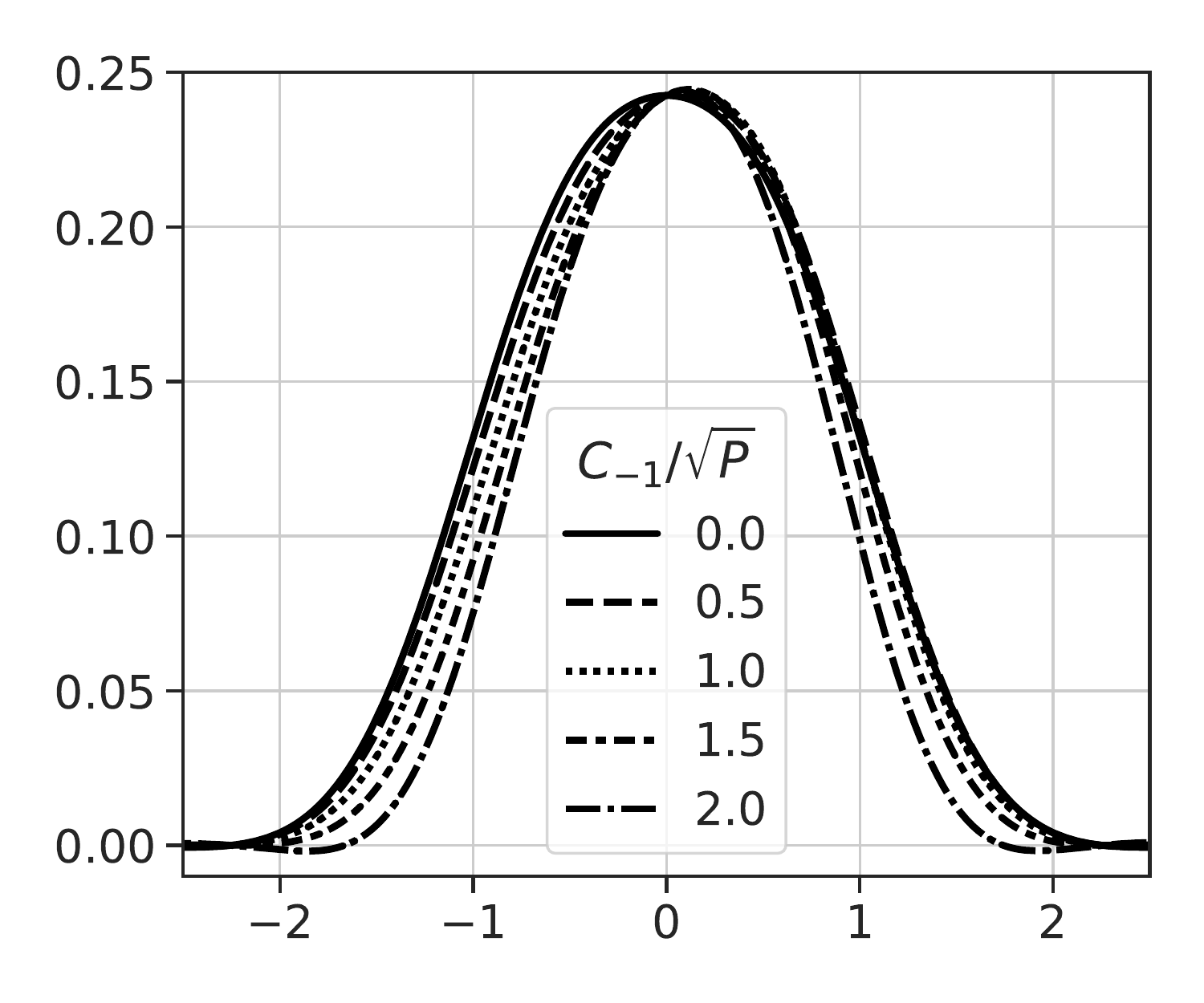} 
    \hfill
    \includegraphics[width=0.45\linewidth]{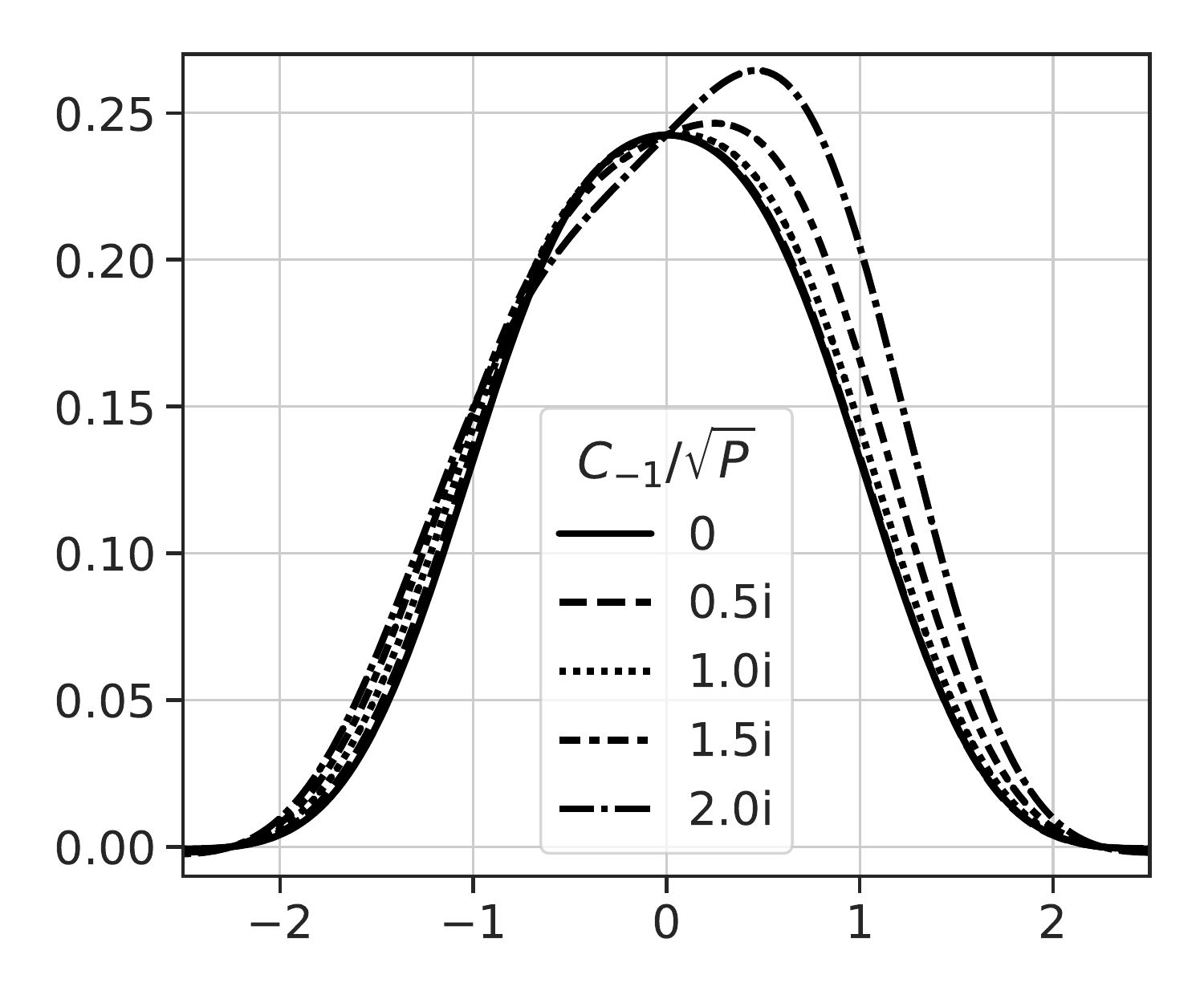} \\
    \includegraphics[width=0.45\linewidth]{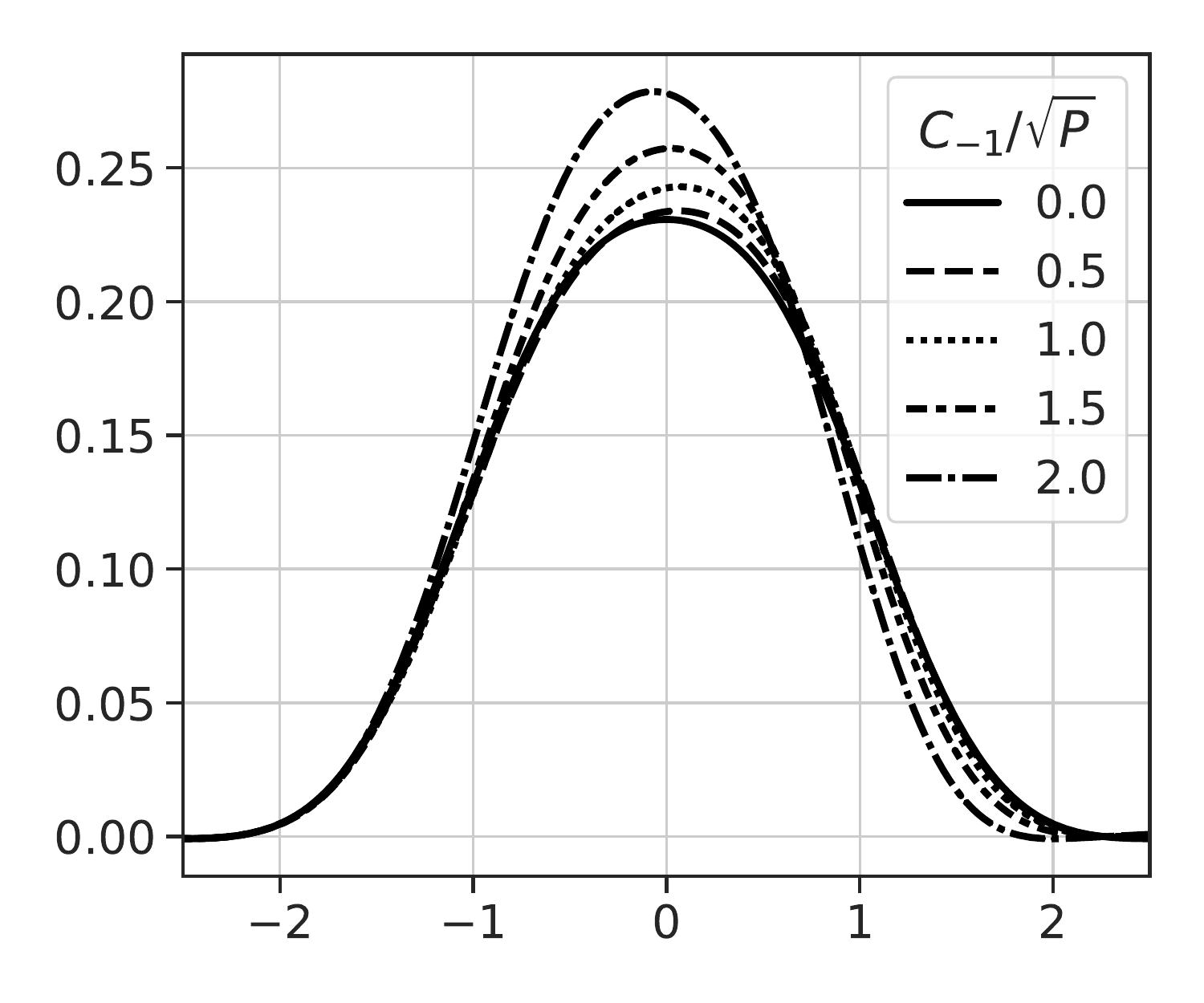}
     \hfill
    \includegraphics[width=0.45\linewidth]{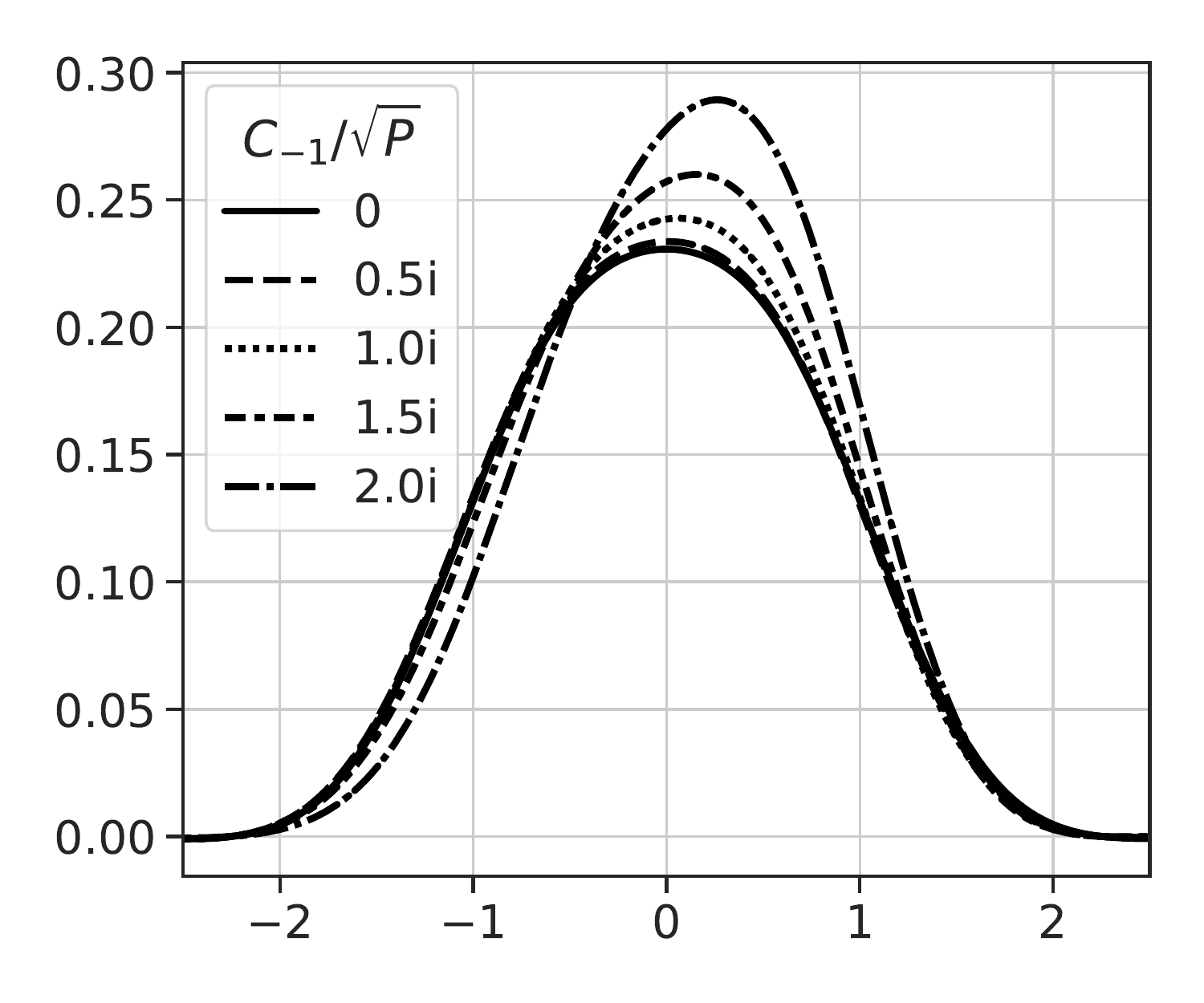}
    \begin{picture}(0,0)
\put(-115,0){(d)} \put(-400,0){(c)} \put(-115,190){(b)}
\put(-400,190){(a)}
\end{picture}
\caption{The PDF $P_{opt}[C_{0}|C_{-1}]$ as a function of real
$C_{0}$ for different values of $C_{-1}/\sqrt{P}$ and
$\tilde{\beta}$. Figs. (a) and (b) correspond to the case
$\tilde{\beta}=1$ for real and imaginary values of
$C_{-1}/\sqrt{P}$, respectively. Figs. (c) and (d) correspond to the
case $\tilde{\beta}=5$ for real and imaginary values of
$C_{-1}/\sqrt{P}$, respectively.}
    \label{Fig4}
\end{figure}

The expression in the big curly brackets in Eq.~\eqref{P[CC2]} is
symmetric in the coefficients $C_i$ and $C_j$.  Since we know the
function $P_{opt}[C_j,C_i]$ the probability $P_{opt}[C_j|C_i]$ can
be easily obtained using Eq.~\eqref{P[CC]General}:
\begin{eqnarray}\label{P[C|C2]}
&&P_{opt}[C_i|C_j]= P_{opt}[C_i]\left\{ 1+(\gamma P L)^2
D^{i,j}\left(\frac{C_i}{\sqrt{P}},\frac{C_j}{\sqrt{P}}\right)
\right\}.
\end{eqnarray}
In Fig.~\ref{Fig4} we plot the function $P_{opt}[C_{0}|C_{-1}]$ for
different values of coefficient $C_{-1}/\sqrt{P}$ (real and
imaginary), nonlinearity parameter $(\gamma L P)^2=0.2$, and
dispersion parameter $\tilde{\beta}$ equals to $1$ and $5$. We plot
the dependence of $P_{opt}[C_{0}|C_{-1}]$ on the dimensionless
variable $C_{0}/\sqrt{P}$, where $P$ is chosen to be equal to the
unity. One can see that the function $P_{opt}[C_{0}|C_{-1}]$ differs
from $P_{opt}[C_{0}]$ and depends on the value $C_{-1}$ essentially.
Also, the PDFs deviate from
the Gaussian distribution greater for larger absolute values of $C_{-1}$. One can see  in
Figs.~\ref{Fig4}(a) and (c) that the PDF reaches the negative value
at the vicinity of $|C_{0}|\sim 2$. The reason for that is the large
chosen nonlinearity parameter $(\gamma L P)^2=0.2$. The negative
value of the PDF has no sense, but it demonstrates the region of
applicability of our approximation. For smaller parameters $\gamma L
P$ or for larger parameter $\tilde{\beta}$ our perturbative result
\eqref{P[CC2]} is valid wherever the function $P_{opt}[C_j,C_i]$ is
not small.
For the large values of $\tilde{\beta}$ it is necessary to consider
not only PDF $P_{opt}[C_j,C_i]$, but others PDFs presented in
Eqs.~\eqref{PoptC1}--\eqref{PoptCM}, since the spreading effects
become significant. If necessary, PDFs
$P_{opt}[C_i|C_{j_1},C_{j_2}]$, \ldots,
$P_{opt}[C_i|C_{j_1},C_{j_2},\ldots, C_{j_{2M}}]$  can be derived
from \eqref{Popt} analogously to Eq.~\eqref{P[C|C2]}. We do not
present these PDFs here because of its cumbersomeness. We attach the
files with coefficients $J_I^{i,j;k,l}$ for different $M$ and
$\tilde{\beta}$ to the possibility to calculate these PDFs.

\section{Conclusion}

In the present paper we develop the method of the calculation of the
conditional probability density function $P[\{\tilde{C}\}|\{{C}\}]$
for the channel describing by the nonlinear Schr\"{o}dinger equation with
additive noise and with nonzero second dispersion coefficient
$\beta$. To illustrate our method we calculate the PDF
$P[\{\tilde{C}\}|\{{C}\}]$ in the leading and next-to-leading order
in the Kerr nonlinearity parameter $\gamma$ and in the leading order
in the parameter $1/\mathrm{SNR}$. To obtain
$P[\{\tilde{C}\}|\{{C}\}]$ we calculated $P[Y(\omega)|X(\omega)]$
using two different approaches. The first approach is based on the
direct calculation of the path-integral, see Eq. \eqref{pathintPYX}.
In the  second approach we calculate the output signal correlators
for the fixed input signal $X(t)$ and then construct the conditional
PDF. Both approaches give the same result \eqref{PYX1}. To take into
account the envelope of the input signal and the detection procedure
of the receiver we integrate the PDF $P[Y(\omega)|X(\omega)]$ over
the redundant degrees of freedom and obtain the conditional PDF
$P[\{\tilde{C}\}|\{{C}\}]$. Using the PDF $P[\{\tilde{C}\}|\{{C}\}]$ we calculate
the mutual information, solve the variational problem, and find the
optimal input signal distribution $P_{opt}[\{C\}]$ in the leading
order in the parameter $1/\mathrm{SNR}$ and in the second order in
the parameter of the Kerr nonlinearity $\gamma$. We demonstrate that
$P_{opt}[\{C\}]$ differs from the Gaussian distribution. Using the
distribution $P_{opt}[\{C\}]$ we calculated the maximal value of the
mutual information  in the leading order in the parameter
$1/\mathrm{SNR}$ and in the second order in the parameter $\gamma$
for the given pulse envelope, average power, and detection
procedure. We demonstrate that the $\gamma^2$-correction to the
mutual information is negative. Its absolute value is maximal  for the zero
dispersion, and it decreases for increasing
dispersion parameter $\tilde{\beta}$, see Fig.~\ref{figPoptI}. We
also prove that the mutual information calculate using the Gaussian
distribution and that calculated with the optimal one coincide in
the $\gamma^2$ order. The difference appears only in the $\gamma^4$
order. It means that the Gaussian distribution of the input signal
is a good approximation of the optimal distribution for the small
nonlinearity parameter. However, for not extremely small
nonlinearity parameter it is necessary to take into account the
exact PDF  $P_{opt}[\{C\}]$. So, we are able to construct the
sequences $\{C\}$ obeying the statistics $P_{opt}[\{C\}]$. In the
Section \ref{SectionConstruction} we propose the method of this
construction using the conditional PDFs
\eqref{PoptC1}--\eqref{PoptCM}. For the channels with the small
correlation length we calculated explicitly $P_{opt}[C_{i}]$ and
$P_{opt}[C_{i}|C_{j}]$, and demonstrate the dependence of the
probability of the subsequent coefficient $C_i$ on the previous one
$C_j$, see Fig.~\ref{Fig4}.

\begin{acknowledgments}
\emph{Acknowledgment}\\
The work of A.V. Reznichenko was supported by the Ministry of Education and Science of the Russian Federation. The work of I.S. Terekhov and E.V. Sedov was supported by the Russian Science Foundation (RSF), grant No. 17-27-30006.
\end{acknowledgments}

\appendix

\section{Details of the conditional PDF $P[\{\tilde{C}\}|\{{C}\}]$ calculation}
\label{AppCondPDFcalc} In the following we present the calculation
of the conditional PDF $P_{d}[Y(\omega)|X(\omega)]$ resulting in the
correlators \eqref{defcorrelatorsC}. In turn, these correlators
enable us to construct the conditional PDF
$P[\{\tilde{C}\}|\{{C}\}]$ in the form \eqref{PtildeCC0}. In this
Appendix we present the explicit expression for the coefficients
$F^{k',k}$, $H^{k',k}$, and $G^{k',k}=\bar{H}^{k',k}$ appearing in
the formula \eqref{PtildeCC0}. Finally, we present the explicit
expression for the normalization factor $\Lambda_c$  in
Eq.~\eqref{PtildeCC0}.


We start our calculation from the representation for the conditional
PDF $P[Y(\omega)|X(\omega)]$ in the form of the path-integral \eqref{pathintPYX}, where the effective action $S[\psi]$ is given by the formula \eqref{action}, see \cite{Terekhov:2014}. The measure ${\cal D} \psi (z,\omega)$ can be
presented in  a specific discretization scheme in such a way that
\begin{eqnarray}
\int {\cal D} Y(\omega) P[Y(\omega)|X(\omega)] = 1,
\end{eqnarray}
where ${\cal D}Y(\omega)=\prod_{k=-\tilde{M}}^{\tilde{M}} d Re
Y(\omega_k) d Im Y(\omega_k)$, here $2\tilde{M}+1$ is the total
number of the discrete points in frequency domain $\tilde{W}$ with
the spacing $2 \pi \delta_{\omega}=\tilde{W}/(2\tilde{M})$. Let us stress, that the bandwidth $\tilde{W}$ is associated with the noise bandwidth, see Eq.~\eqref{noisecorrelatort}. For the discretization in $z$ coordinate with $N$ points with spacing
$\delta_{z}=L/(N-1)$ one has
\begin{eqnarray} \label{SMmeasureomega}
&&{\cal D} {\psi (z,\omega)}=\Big( \frac{\delta_{\omega}}{\delta_{z}
\pi
Q}\Big)^{2\tilde{M}+1}\prod^{\tilde{M}}_{j'=-\tilde{M}}\prod^{N-1}_{k=1}\Big\{
\frac{\delta_{\omega}}{\delta_{z} \pi Q}\,\,
dRe\psi(z_k,{\omega}_{j'})\,dIm\psi(z_k, {\omega}_{j'})\Big\}.
\end{eqnarray}

\subsection{Short notations}
First of all, for brevity sake let us introduce the following
notations:
\begin{eqnarray}\label{parenthesis1}
({\omega_1},{\omega_2};{\omega_3},{\omega_4})_{z}=\frac{e^{i \beta z
(\omega^2_1+\omega^2_2-\omega^2_3-\omega^2_4)}}{(2\pi)^2}
\delta(\omega_1+\omega_2-\omega_3-\omega_4),
\end{eqnarray}
and for any compactly supported functions of frequencies
$F^{(1)}({\omega_1})$, $F^{(2)}({\omega_2})$, and complex conjugated
functions $\bar{F}^{(3)}({\omega_3})$, $\bar{F}^{(4)} ({\omega_4})$
we use  notations:
\begin{eqnarray}\label{parenthesis2}
(F^{(1)},F^{(2)};\bar{F}^{(3)},\bar{F}^{(4)})_z&=& \int_{\tilde{W}} d \omega_1 d \omega_2 d \omega_3 d \omega_4 ({\omega_1},{\omega_2};{\omega_3},{\omega_4})_{z} F^{(1)}({\omega_1})F^{(2)}({\omega_2})\bar{F}^{(3)}({\omega_3})\bar{F}^{(4)}({\omega_4}),\nonumber\\
(F^{(1)},F^{(2)};\bar{F}^{(3)},\omega_4)_z&=& \int_{\tilde{W}} d \omega_1 d \omega_2 d \omega_3  ({\omega_1},{\omega_2};{\omega_3},{\omega_4})_{z} F^{(1)}({\omega_1})F^{(2)}({\omega_2})\bar{F}^{(3)}({\omega_3}),\nonumber\\
(F^{(1)},F^{(2)};\omega_3,\omega_4)_z&=& \int_{\tilde{W}} d \omega_1 d \omega_2   ({\omega_1},{\omega_2};{\omega_3},{\omega_4})_{z} F^{(1)}({\omega_1})F^{(2)}({\omega_2}),\nonumber\\
(F^{(1)},\omega_2;\omega_3,\omega_4)_z&=& \int_{\tilde{W}} d
\omega_1 ({\omega_1},{\omega_2};{\omega_3},{\omega_4})_{z}
F^{(1)}({\omega_1}).
\end{eqnarray}
It is obvious that this notations have the properties:
\begin{eqnarray}\label{parenthesis3}
&&({\omega_1},{\omega_2};{\omega_3},{\omega_4})_{z}=({\omega_2},{\omega_1};{\omega_3},{\omega_4})_{z}=({\omega_1},{\omega_2};{\omega_4},{\omega_3})_{z},\nonumber
\\&&
\overline{(F^{(1)},F^{(2)};\bar{F}^{(3)},\omega)_z}=(\omega,F^{(3)};\bar{F}^{(1)},\bar{F}^{(2)})_z.
\end{eqnarray}

\subsection{The solution $\Phi(z,\omega)$ of the NLSE with zero noise. The Jacobian $\left|\frac{\partial
(\tilde{C}^{(0)},\bar{\tilde{C}}^{(0)})} {\partial
(C,\bar{C})}\right|$.}

The solution $\Phi(z,\omega)$ of the NLSE~(\ref{startingCannelEqt}) in the frequency domain with zero noise and with the input
condition $\Phi(0,\omega)=X(\omega)$ plays an important role in our
consideration. Here we present the perturbative result for this
function for the small Kerr nonlinearity $\gamma$ ($\gamma L P \ll 1$).  The solution $\Phi(z, {\omega})$ reads in the first and
in the second order in the Kerr nonlinearity parameter $\gamma$:
\begin{eqnarray} \label{Phizo}
&&{\Phi}(z,{\omega})= e^{i \beta z \omega^2} \Big\{X({\omega})+i
\gamma \int^z_0 dz' (X,X;\bar{X},\omega)_{z'}-\gamma^2 \int^{z}_0
dz_1 \int^{z_1}_0 dz_2 \int_{\tilde{W}} d\omega_a \Big[2
(\omega_a,X;\bar{X},\omega)_{z_1} (X,X;\omega_a,\bar{X})_{z_2} -
\nonumber \\&& (X,X;\omega_a,\omega)_{z_1}
(\omega_a,X;\bar{X},\bar{X})_{z_2}\Big] \Big\},
\end{eqnarray}
where we have used the notations \eqref{parenthesis2}. The function
$X(\omega)$ appearing in the representation \eqref{Phizo} is the
Fourier integral of the input signal, see Eq.~\eqref{Xtmodelg}:
\begin{eqnarray}
X({\omega})\equiv \int^{\infty}_{-\infty}dt X(t) e^{i \omega t} =
\sum^{M}_{k=-M} C_{k} \, s^{(k)}({\omega}), \qquad
s^{(k)}({\omega})=s({\omega}) e^{ik\omega T_0}.
\end{eqnarray}
Note that $s({\omega})$ is not zero only within the frequency domain $W$:
$|\omega|<W/2$.

Let us introduce the coefficients $\tilde{C}^{(0)}_k$ obtained from
the function ${\Phi}(z,{\omega})$ according to the procedure
Eq.~\eqref{tildeCk}
\begin{eqnarray}\label{ctildezero}
&& \tilde{C}^{(0)}_k= \frac{1}{T_0}\int_{W_d} \frac{d
\omega}{2\pi}e^{-i\beta L
\omega^2}{\Phi}(L,{\omega})\bar{s}_{\omega}^{(k)}={C}_k+ i{\gamma L}
{C}_{k_1}{C}_{k_2}\bar{{C}}_{k_3}a^{k_1,k_2;{k_3},{k}}_{1}
-\nonumber \\&&  {\gamma^2 L^2}
{C}_{m_1}{C}_{m_2}{C}_{m_3}\bar{{C}}_{m_4}\,\bar{{C}}_{m_5}a^{m_1,m_2,m_3;\,{m_4},{m_5},{k}}_2,
\end{eqnarray}
where hereinafter  the summation is assumed over all repeated
indexes from $-M$ to $M$, and in Eq.~(\ref{ctildezero}) we
introduce the dimensionless coefficients:
\begin{eqnarray}\label{Akoeff}
a^{k_1,k_2;{k_3},{k_4}}_1={\frac{1}{2\pi T_0}}\int^L_ 0 \frac{dz}{L}
(s^{(k_1)},s^{(k_2)};\bar{s}^{(k_3)},\bar{s}^{(k_4)})_z,
\end{eqnarray}
\begin{eqnarray}\label{Bkoeff}
&&a^{m_1,m_2,m_3;\,{m_4},{m_5},{m_6}}_2= \frac{1}{2\pi T_0}\int^L_ 0
\frac{dz_1}{L} \int^L_ 0 \frac{dz_2}{L} \int_{\tilde{W}} d\omega_a
\Bigg[2 \theta(z_1-z_2)
(\omega_a,s^{(m_3)};\bar{s}^{(m_5)},\bar{s}^{(m_6)})_{z_1}(s^{(m_1)},s^{(m_2)};\omega_a,\bar{s}^{(m_4)})_{z_2}
- \nonumber \\&&   \theta(z_2-z_1)
(\omega_a,s^{(m_3)};\bar{s}^{(m_4)},\bar{s}^{(m_5)})_{z_1}
(s^{(m_1)},s^{(m_2)};\omega_a,\bar{s}^{(m_6)})_{z_2} \Bigg].
\end{eqnarray}
The meaning of the coefficients $\tilde{C}^{(0)}_k$ is as follows:
for the nonlinear channel with zero noise the coefficients
$\tilde{C}^{(0)}_k$ are the recovered coefficients $\{{C} \}$ on the
base of the output signal ${\Phi}(L,t)$ according to our procedure
\eqref{tildeCk}. We can restore all $\{{C} \}$ unequivocally in the
perturbation theory, if we know all $\tilde{C}^{(0)}_k$.

The Jacobian determinant $\left|\frac{\partial
(\tilde{C}^{(0)},\bar{\tilde{C}}^{(0)})} {\partial
(C,\bar{C})}\right|$ plays an important role in the  output entropy
calculation, see Eq.~\eqref{HtildeC}.

Therefore to find the output entropy we should calculate the
logarithm of the determinant in Eq.~(\ref{HtildeC}). To calculate it
we use  Eq.~(\ref{ctildezero}), and with $\gamma^2$ accuracy we
obtain
\begin{eqnarray}\label{Jacobian2}
&& \log  \left|\frac{\partial
(\tilde{C}^{(0)},\bar{\tilde{C}}^{(0)})} {\partial
(C,\bar{C})}\right|= {\gamma^2 L^2}
{C}_{s_1}{C}_{s_2}\bar{{C}}_{s_3}\,\,\bar{{C}}_{s_4}  J^{s_1,s_2;
{s_3},{s_4}},
\end{eqnarray}
where dimensionless coefficients $J^{s_1,s_2;{s_3},{s_4}}$ have the
form:
\begin{eqnarray}\label{JJacobian}
&&J^{s_1,s_2;{s_3},{s_4}}= \hat{\cal S}_{12,34}\Big[\Bigg( 2
a_1^{r',s_1;{s_3},{r}}a_1^{r,s_2;{s_4}, {r'}}- \frac{1}{2}
a_1^{s_1,s_2;{r},{r'}}a_1^{r',r;{s_3},{s_4}}- 2
a_2^{r,s_1,s_2;\,{s_3},{s_4},{r}}-
a_2^{s_1,s_2,r;\,{s_3},{s_4},{r}}\Bigg)+\nonumber \\&&
\overline{\Bigg(s_1 \leftrightarrow s_3, s_2 \leftrightarrow
s_4\Bigg)}\Big],
\end{eqnarray}
here $\hat{\cal S}_{12,34}$ is the symmetrization operator that makes
the left hand side to be symmetric under substitutions $s_1
\leftrightarrow s_2$ and $s_3 \leftrightarrow s_4$.  We represent
the convenient combination of coefficients (\ref{Bkoeff}):
\begin{eqnarray}
a^{m_1,m_2,m_3;\,{m_4},{m_5},{m_6}}_2= 2
A^{m_1,m_2,m_3;\,{m_4},{m_5},{m_6}}_2 -
\left(A^{m_4,m_5,m_6;\,{m_3},{m_1},{m_2}\,}_2\right)^*,
\end{eqnarray}
where $A^{m_1,m_2,m_3;\,{m_4},{m_5},{m_6}}_2 $ is the ninefold
integral
\begin{eqnarray}\label{A2form01}
&&A^{m_1,m_2,m_3;\,{m_4},{m_5},{m_6}}_2=\frac{1}{2\pi T_0}\int^L_ 0
\frac{dz_1}{L} \int^{z_1}_ 0 \frac{dz_2}{L} \int_{\tilde{W}}
d\omega_a
(\omega_a,s^{(m_3)};\bar{s}^{(m_5)},\bar{s}^{(m_6)})_{z_1}(s^{(m_1)},s^{(m_2)};
\omega_a,\bar{s}^{(m_4)})_{z_2}.
\end{eqnarray}

After simplifying \eqref{JJacobian}, we present coefficients $J^{s_1,s_2;{s_3},{s_4}}$ in the form
\begin{eqnarray}\label{JJacobian2}
&&J^{s_1,s_2;{s_3},{s_4}}= \hat{\cal S}_{12,34}\Bigg( 4
a_1^{r',s_1;{s_3},{r}}a_1^{r,s_2;{s_4}, {r'}}-
a_1^{s_1,s_2;{r},{r'}}a_1^{r',r;{s_3},{s_4}}- \Big[ 4
A_2^{r,s_1,s_2;\,{s_3},{s_4},{r}}-
A_2^{s_1,s_2,r;\,{r},{s_3},{s_4}}\Big]\nonumber
\\&&-\overline{\Big[4
A_2^{r,s_3,s_4;\,{s_1},{s_2},{r}}-
A_2^{s_3,s_4,r;\,{r},{s_1},{s_2}}\Big]}\Bigg),
\end{eqnarray}
where $\hat{\cal S}_{12,34}$ is the symmetrization operator under
the changes $s_1 \leftrightarrow s_2$ and  $s_3 \leftrightarrow
s_4$. The explicit result of the symmetrization reads:
\begin{eqnarray}\label{JJacobian4}
&&J^{s_1,s_2;{s_3},{s_4}}=2 a_1^{r',s_1;s_3,r}a_1^{r,s_2;s_4,r'}+2
a_1^{r',s_2;s_3,r}a_1^{r,s_1;s_4,r'}-a_1^{s_1,s_2;r,r'}a_1^{r,r';s_3,s_4}+\nonumber
\\&&\Big[A_2^{s_1,s_2,r;\,r,s_3,s_4}-A_2^{r,s_1,s_2;\,s_3,s_4,r}-A_2^{r,s_2,s_1;\,s_3,s_4,r}-
A_2^{r,s_1,s_2;\,s_4,s_3,r}-A_2^{r,s_2,s_1;\,s_4,s_3,r}\Big]+\overline{\Big[s_1
\leftrightarrow s_3, s_2 \leftrightarrow s_4\Big]}.
\end{eqnarray}
The methods of the numerical calculation of these coefficients
$a^{k_1,k_2;{k_3},{k_4}}_1$ and $A_2^{s_1,s_2,s_3;\,s_4,s_5,s_6}$ for the sinc envelope \eqref{Sincft} are presented in the Appendix \ref{NumCalcSection}.

Finally, let us note here that the perturbation theory in the
nonlinearity parameter $\gamma$ allows us to estimate the spectral broadening
of the signal which propagation is governed by the NSLE with zero
noise. We can define the effective spectral bandwidth of the input
signal $X(t)$ as follows
\begin{eqnarray}
W^2_i={\int \frac{d \omega}{2 \pi} \omega^2 |X({\omega})|^2}/{\int
\frac{d \omega}{2 \pi} |X({\omega})|^2}.
\end{eqnarray}
For the sinc envelope \eqref{Sincft} we have $W^2_i=\frac{W^2}{12}$.
In the same manner we can define the effective bandwidth of the
output noiseless signal $\Phi(L, t)$:
\begin{eqnarray}
W^2_f={\int \frac{d \omega}{2 \pi} \omega^2
|{\Phi}(L,{\omega})|^2}/{\int \frac{d \omega}{2 \pi}
|{\Phi}(L,{\omega})|^2},
\end{eqnarray}
where ${\Phi}(z,{\omega})$ is given in Eq.~\eqref{Phizo} in
perturbation expansion. From this representation it is easy to find
that
\begin{eqnarray}
W_f=W_i\left(1+\frac{\gamma L}{4 (\beta L W^2_i) \int dt
|X(t)|^2}\left\{ \int dt |X(t)|^4 - \int dt |\Phi_{\gamma=0}(L,t)|^4
\right\}+ {\cal O}(\gamma^2)\right),
\end{eqnarray}
where $\Phi_{\gamma=0}(z,t)$ is the solution of the linear noiseless
Schr\"{o}dinger equation (i.e. equation \eqref{startingCannelEqt}
with $\gamma=0$ and $\eta=0$). It is especially simple in the
frequency domain: $\Phi_{\gamma=0}(z,\omega)=e^{i \beta z \omega^2}
X({\omega})$, and in the time domain it reads
\begin{eqnarray}
\Phi_{\gamma=0}(z,t)= \int \frac{d \omega}{2 \pi} X({\omega}) e^{i
\beta z \omega^2- i\omega t}=\frac{\theta(z)}{\sqrt{4 \pi \beta z}}
\int dt' X(t') e^{i \frac{\pi}{4}-i \frac{(t-t')^2}{4 \beta z}}.
\end{eqnarray}

\subsection{Integration over fields $\psi (z,\omega)$ and over $Y(\omega)$ with $|\omega|>W_d$ in path-integral \eqref{pathintPYX}}
To calculate the path-integral (\ref{pathintPYX}) we perform the
change of variables from $\psi (z,\omega)$ to $\phi(z,\omega)$,
where
\begin{eqnarray}\label{phivar}
\psi (z,\omega)=e^{i \beta z \omega^2} \phi(z,\omega) +
\Phi(z,\omega) +\frac{z}{L}B(\omega),
\end{eqnarray}
where $\Phi(z,\omega)$ is the solution of the
Eq.~(\ref{startingCannelEqt}) with zero noise and with the input
condition $\Phi(0,\omega)=X(\omega)$, and
$B(\omega)=Y(\omega)-\Phi(L,\omega)$. The integration over new
variables $\phi(z,\omega) $ is performed with the boundary
conditions $\phi(0,\omega)=\phi(L,\omega)=0$. Since we calculate
$P[Y(\omega)|X(\omega)]$ in the leading order in the parameter
$1/\mathrm{SNR}$, after substitution of the function  $\psi
(z,\omega)$ in the form (\ref{phivar}) to Eq.~(\ref{action}) we
should retain only terms  of the orders of $\phi^0(z,\omega)$,
$\phi^1(z,\omega)$, and $\phi^2(z,\omega)$. Then we expand the
exponent in Eq.~(\ref{pathintPYX}) in the parameter $\gamma$ up to
$\gamma^2$ terms, and perform the Gaussian integration over $\phi
(z,\omega)$.

The result of Gaussian integration over fields $\phi(z,\omega)$
depends on the function $X(\omega)$ and $B(\omega)$, where the
frequency $\omega \in [-\tilde{W}/2, \tilde{W}/2]$.  According to Eq.~\eqref{Pdintegration}, to obtain the conditional PDF
$P_{d}[Y_d(\omega)|X(\omega)]$ we should integrate over
$B(\omega)=Y(\omega)-\Phi(L,\omega)$ for $\omega \in [-\tilde{W}/2,
-{W}_d/2] \cup [{W}_d/2, \tilde{W}/2]$ and we arrive at the result:
\begin{eqnarray}\label{PYX1}
&& P_{d}[Y_d|X]= \Lambda_d\exp\Big\{-\frac{1}{Q L}
\int_{W_d}\int_{W_d}\frac{d \omega_1d \omega_2}{(2
\pi)^2}\Big(\delta\tilde{Y}({\omega_1})F(\omega_1,\omega_2)
\overline{\delta\tilde{Y}}({\omega_2})+\delta
\tilde{Y}({\omega_1})G(\omega_1,\omega_2)
{\delta\tilde{Y}({\omega_2})}+\nonumber \\&& \overline{\delta
\tilde{Y}({\omega_1})}H(\omega_1,\omega_2)\overline{\delta\tilde{Y}({\omega_2})}\Big)\Big\}.
\end{eqnarray}
One can see that frequency variables in Eq.~(\ref{PYX1}) are from
the interval $\omega \in [-W_d/2, W_d/2]$, rather than
$[-\tilde{W}/2, \tilde{W}/2]$ as in Eq.~(\ref{pathintPYX}).  In
Eq.~(\ref{PYX1}) we have introduced the following notations:
\begin{eqnarray}
\delta\tilde{Y}({\omega})=e^{-i \beta L \omega^2}\Big(
Y_{d}(\omega)-  {\cal Y}_{d}(\omega) \Big),
\end{eqnarray}
where the quantity ${\cal Y}_{d}(\omega)$, see Eq. \eqref{tildeYav} below, reads
\begin{eqnarray}\label{tildeYdav}
&& {\cal Y}_{d}(\omega)={\Phi}(L,{\omega})+i \frac{Q L \tilde{W}}{2
\pi} L \gamma {\Phi}(L,{\omega}) -\nonumber \\&& 4 \pi Q \gamma^2 L
e^{i \beta L \omega^2} \int^{L}_0 dz_1 \int^{L}_0 dz_2
\int_{\tilde{W}} d\omega_a \int_{\tilde{W}} d\omega_b
\theta(z_1-z_2) \frac{z_2}{L} (\omega_a,
\omega_b;\overline{X},\omega)_{z_1}(X,X;\omega_a, \omega_b)_{z_2},
\end{eqnarray}
where
$F(\omega_1,\omega_2)=F_0(\omega_1,\omega_2)+F_1(\omega_1,\omega_2)+F_2(\omega_1,\omega_2)$,
$H(\omega_1,\omega_2)=H_0(\omega_1,\omega_2)+H_1(\omega_1,\omega_2)+H_2(\omega_1,\omega_2)$,
and $G(\omega_1,\omega_2)=\bar{H}(\omega_1,\omega_2)$, here
\begin{eqnarray}\label{Fresult}
&&F_{0}(\omega_1,\omega_2)=2 \pi \delta(\omega_1-\omega_2), \qquad
F_{1}(\omega_1,\omega_2)=0;\nonumber \\&& F_{2}(\omega_1,\omega_2)=
4 \pi \gamma^2\int^L_0 dz_1\int^L_0 dz_2 \Big( 2 \frac{z_1
z_2}{L^2}\int_{W_d}d\omega_a
-\frac{\min(z_1,z_2)}{L}\int_{\tilde{W}}d\omega_a\Big)
(\omega_a,\omega_1;\bar{X},\bar{X})_{z_1}(X,X;\omega_a,\omega_2)_{z_2},
\end{eqnarray}
\begin{eqnarray}\label{Hresult}
&&H_{0}(\omega_1,\omega_2)=0, \qquad H_{1}(\omega_1,\omega_2)=-2\pi
i \gamma \int^L_0 dz \frac{z}{L} (X,X;\omega_1,\omega_2)_z
;\nonumber \\&& H_{2}(\omega_1,\omega_2)= 4 \pi \gamma^2\int^L_0
dz_1 \int^{L}_0 dz_2  \theta(z_1-z_2) \int_{\tilde{W}} d \omega_a
\Bigg[\frac{z_1}{L} (\omega_a,X;\omega_1,\omega_2)_{z_1}
(X,X;\omega_a,\bar{X})_{z_2}+ \nonumber \\&&\!\!\!\!\! \frac{z_2}{L}
(\omega_a,X;\bar{X},\omega_1)_{z_1}
(X,X;\omega_a,\omega_2)_{z_2}+\frac{z_2
}{L}(\omega_a,X;\bar{X},\omega_2)_{z_1}
(X,X;\omega_a,\omega_1)_{z_2}\Bigg].
\end{eqnarray}

The normalization factor $\Lambda_d$ in Eq.~(\ref{PYX1}) reads
\begin{eqnarray}\label{Lambdaf}
\Lambda_d= \left(\frac{\delta_{\omega}}{\pi Q
L}\right)^{2M_d+1}\left(1+\lambda_2 \right),
\end{eqnarray}
where $\delta_{\omega}={W_d}/(2{M_d})$ is the frequency grid spacing
in the discretization of the path-integral in the frequency domain, $2M_d+1$ is the frequency discretization number. It means that the integrals over
frequencies in Eq.~(\ref{PYX1}) are understood as the sums:
$\int_{W_d} \frac{d \omega}{2\pi} g(\omega) = \delta_{\omega}
\sum_{k=-M_d}^{M_d} g(\omega_k)$, where $\omega_k= k
\delta_{\omega}$ and $g(\omega)$ is any function. The explicit result for
$\lambda_2$ reads
\begin{eqnarray}\label{lambda2-1}
\lambda_2&=&2 \gamma^2 \int^L_0 dz_1 \int^L_0 dz_2
\frac{z_1z_2}{L^2} \int_{W_d}d\omega_a\int_{W_d}d\omega_b
(\omega_a,\omega_b;\bar{X},\bar{X})_{z_1}(X,X;\omega_a,\omega_b)_{z_2}-\nonumber
\\&& 2 \gamma^2 \int^L_0 dz_1 \int^L_0 dz_2
\frac{\min(z_1,z_2)}{L}\int_{W_d}d\omega_a \int_{\tilde{W}}d\omega_b
(\omega_a,\omega_b;\bar{X},\bar{X})_{z_1}(X,X;\omega_a,\omega_b)_{z_2}.
\end{eqnarray}

Note that in Eq.~(\ref{PYX1}) the exponent is understood as a series
with the retained terms up to $\gamma^2$ only. If we set $\gamma=0$
in Eq.~(\ref{PYX1}) we result in the Gaussian distribution:
\begin{eqnarray}\label{gaussianlimit}
\lim _{\gamma \to 0} P_{d}[Y_d|X]=\left(\frac{\delta_{\omega}}{\pi Q
L}\right)^{2M_d+1} \exp\left[-\frac{1}{Q L} \int_{W_d} \frac{d
\omega}{2\pi} \left| Y_d(\omega)-e^{i \beta L \omega^2}X(\omega)
\right|^2 \right].
\end{eqnarray}
One can check that in the limit $Q \to 0$ the conditional PDF
(\ref{PYX1}) reduces to the Dirac delta-function:
\begin{eqnarray}\label{limitQ}
\lim _{Q \to 0} P_{d}[Y_d|X]=\prod_{k=-M_d}^{M_d}
\delta(Y_d(\omega_k) - \Phi(L,\omega_k)),
\end{eqnarray}
where $\omega_k=k \delta_{\omega}$. Also the conditional PDF
(\ref{PYX1}) obeys the normalization condition:
\begin{eqnarray}
\int {\cal D}Y_{d} P_{d}[Y_d|X] =1,
\end{eqnarray}
where the functional measure reads ${\cal
D}Y_{d}=\prod_{k=-M_d}^{M_d} d Re Y_{d}(\omega_k) d Im
Y_{d}(\omega_k)$. Note that the quantity ${\cal Y}_{d}(\omega)$
depends only on the input signal $X(t)$ and has the meaning of the
average of the output signal over the distribution (\ref{PYX1}):
\begin{eqnarray}\label{tildeYav}
&&{\cal Y}_{d}(\omega) = \langle Y_d(\omega) \rangle \equiv  \int
{\cal D} Y_d P_{d}[Y_d|X] {Y}_{d}(\omega).
\end{eqnarray}
Let us emphasize that the terms proportional to the noise power $Q$
in the expression (\ref{tildeYdav}) for ${\cal Y}_{d}(\omega) $
contain the whole noise bandwidth $\tilde{W}$ rather than the
detector bandwidth $W_d$. This is the consequence of the effects of
nonlinearity: due to the Kerr nonlinearity the signal mixes with the
noise in the whole frequency interval.

\subsection{The correlators of $\delta \tilde{C}_k$ and the construction of the PDF $P[\{\tilde{C}\}|\{C\}]$.}

Now we can proceed to the calculation of the correlators of the
coefficients $\tilde{C}_k$, see Eq.~\eqref{tildeCk}. Substituting Eq.~(\ref{PYX1}),
Eq.~(\ref{tildeCk}) to the correlator expression~(\ref{defcorrelatorsC}) and performing integration over $Y_d$ we obtain the following correlator in Kerr
nonlinearity expansion:
\begin{eqnarray}\label{avC}
&&\langle \tilde{C}_k\rangle=\tilde{C}^{(0)}_k+i \frac{Q L
\tilde{W}}{2\pi}\gamma L \tilde{C}^{(0)}_k-\nonumber \\&& \frac{2 Q
L \gamma^2}{T_0} \int^{L}_0 dz_1 \int^{L}_0 dz_2 \int_{\tilde{W}}
d\omega_a \int_{\tilde{W}} d\omega_b \theta(z_1-z_2) \frac{z_2}{L}
(\omega_a,
\omega_b;\bar{s}^{(k_3)},\bar{s}^{(k)})_{z_1}(s^{(k_1)},s^{(k_2)};\omega_a,
\omega_b)_{z_2}\,C_{k_1}C_{k_2}\bar{C}_{k_3},
\end{eqnarray}
where coefficients $\tilde{C}^{(0)}_k$ are defined in Eq.~\eqref{ctildezero} through the coefficients $\{{C}\}$ of the input signal.

For further calculations it is convenient to  introduce the quantity
$\delta \tilde{C}_k$ in the form $\delta \tilde{C}_k=\tilde{C}_k -
\langle \tilde{C}_k\rangle$. This difference is of order of
$\sqrt{Q}$. Therefore to construct the conditional PDF
$P[\{\tilde{C}\}|\{C\}]$ in the leading order in parameter
$1/\mathrm{SNR} \propto Q$ it is sufficient to calculate two
correlators $\langle \delta \tilde{C}_m \delta \tilde{C}_k \rangle$
and $\langle \delta {\tilde{C}}_m \delta \bar{\tilde{C}}_k \rangle$.
After straightforward but cumbersome calculations we arrive at
following result:
\begin{eqnarray} \label{dCdC}
&&\langle \delta \tilde{C}_m \delta \tilde{C}_k \rangle= i \frac{Q L
\gamma}{\pi T_0^2} \int^L_0 dz
\frac{z}{L}(X,X;\bar{s}^{(m)},\bar{s}^{(k)})_{z} - 2  \frac{Q L
\gamma^2 }{\pi T_0^2} \int^L_0 dz_1 \int^{L}_0 dz_2  \theta(z_1-z_2)
\int_{\tilde{W}} d \omega_a \Bigg[ \nonumber \\&& \frac{z_2}{L}
(\omega_a,X;\bar{X},\bar{s}^{(m)})_{z_1}
(X,X;\omega_a,\bar{s}^{(k)})_{z_2}+\frac{z_2}{L}
(\omega_a,X;\bar{X},\bar{s}^{(k)})_{z_1}
(X,X;\omega_a,\bar{s}^{(m)})_{z_2}+\nonumber \\&& \frac{z_1}{L}
(\omega_a,X;\bar{s}^{(m)},\bar{s}^{(k)})_{z_1}
(X,X;\omega_a,\bar{X})_{z_2}\Bigg].
\end{eqnarray}
\begin{eqnarray} \label{dCbardC}
\langle \delta {\tilde{C}}_m \delta \bar{\tilde{C}}_k \rangle=
\frac{Q L}{T_0} \delta_{m k}+  \frac{Q \gamma^2}{\pi T_0^2}
\int^L_0 dz_1 \int^{L}_0  dz_2 \min(z_1,z_2)\int_{\tilde{W}} d
\omega_a
(\omega_a,{s}^{(k)};\bar{X},\bar{X})_{z_1}(X,X;\omega_a,\bar{s}^{(m)})_{z_2}.
\end{eqnarray}
Using the correlators (\ref{avC}), (\ref{dCdC}), and (\ref{dCbardC})
we can calculate any correlator $\langle \tilde{C}_{k_1}\ldots
\bar{\tilde{C}}_{k_N} \rangle$ in the leading ($Q^0$) and
next-to-leading ($Q^1$) order in parameter $Q$. We construct the
distribution $P[\{\tilde{C}\}|\{{C}\}]$ which reproduces all these
correlators in the leading and next-to-leading  order in parameter
$Q$:
\begin{eqnarray}\label{PtildeCC}
P[\{\tilde{C}\}|\{C\}]=\Lambda_c \exp\Big\{-\frac{T_0}{Q L}
\sum_{k,k'=-M}^{M}\Big[ \delta \tilde{C}_{k'} F^{k',k}
\overline{\delta \tilde{C}_k}+ \delta \tilde{C}_{k'} G^{k',k}
{\delta \tilde{C}_k} + \overline{\delta \tilde{C}_{k'}} H^{k',k}
\overline{\delta \tilde{C}_k}\Big] \Big\},
\end{eqnarray}
here $F^{k',k}=\bar{F}^{k,k'}=\delta^{k',k}+F^{k',k}_{2}$,
$H^{k',k}=H^{k',k}_{1}+H^{k',k}_{2}$, $G^{k',k}=\bar{H}^{k',k}=G^{k',k}_1+G^{k',k}_2$ are
dimensionless coefficients with $k, k'=-M, \ldots ,M$. The
subindexes $1$ and $2$ indicate terms proportional to $\gamma^1$ and
$\gamma^2$, respectively. These quantities  depend on the input
signal via coefficients $C_k$. We have
\begin{eqnarray}\label{Hkm}
&&H^{k,m}=H^{m,k}=-\frac{T_0}{2 Q L }\langle \delta\tilde{C}_k \,
\delta\tilde{C}_m \rangle ,
\end{eqnarray}
therefore using Eq.~(\ref{dCdC}) we obtain explicit expressions for
$H^{k',k}_{1}$ and $H^{k',k}_{2}$:
\begin{eqnarray}\label{H12}
&&H^{m,k}_{1}= -i{\gamma L} C_{k_1} C_{k_2} \frac{1}{2 \pi
T_0}\int^{L}_0 \frac{dz}{L}
\frac{z}{L}(s^{(k_1)},s^{(k_2)};\bar{s}^{(k)},\bar{s}^{(m)})_z,
\\ && H^{m,k}_{2}= \gamma^2 L^2 C_{k_1}
C_{k_2}C_{k_3}\bar{C}_{k_4} \frac{1}{\pi
T_0}\int^{L}_{0}\frac{dz_1}{L} \int^{z_1}_{0} \frac{dz_2}{L}
\int_{\tilde{W}}d\omega_a\Bigg[\frac{z_2}{L}(\omega_a,s^{(k_3)};\bar{s}^{(k_4)},\bar{s}^{(m)})_{z_1}(s^{(k_1)},s^{(k_2)};\omega_a,\bar{s}^{(k)})_{z_2}+
\nonumber \\&&
\frac{z_2}{L}(\omega_a,s^{(k_3)};\bar{s}^{(k_4)},\bar{s}^{(k)})_{z_1}(s^{(k_1)},s^{(k_2)};\omega_a,\bar{s}^{(m)})_{z_2}+\frac{z_1}{L}(\omega_a,
s^{(k_3)};\bar{s}^{(m)},\bar{s}^{(k)})_{z_1}(s^{(k_1)},s^{(k_2)};\omega_a,\bar{s}^{(k_4)})_{z_2}
\Bigg],
\end{eqnarray}
\begin{eqnarray}\label{Fcorrections}
&&F^{k',k}_{2}=4 G^{k',m}_{1}H^{m,k}_{1}-\gamma^2\frac{T_0}{2 Q L}
\frac{\partial^2}{\partial \gamma^2}\langle
\delta\bar{\tilde{C}}_{k'} \, \delta\tilde{C}_k
\rangle\Big|_{\gamma=0} =\nonumber \\&&\gamma^2 L^2
{C}_{k_1}{C}_{k_2}\bar{{C}}_{k_3}\,\,\bar{{C}}_{k_4} \frac{1}{(2 \pi
T_0)^2}\int^{L}_0 \frac{dz_1}{L}\int^{L}_0\frac{ dz_2}{L}\Bigg[
 4 \frac{z_1 z_2}{L^2}
(s^{(k_1)},s^{(k_2)};\bar{s}^{(k)},\bar{s}^{(r')})_{z_1}
(s^{(r')},s^{(k')};\bar{s}^{(k_3)},\bar{s}^{(k_4)})_{z_2} -
\nonumber \\&& 4\pi T_0 \frac{\min(z_1,z_2)}{L}\int_{\tilde{W}} d
\omega_a (s^{(k_1)},s^{(k_2)};\omega_a,
\bar{s}^{(k)})_{z_1}(\omega_a,s^{(k')};\bar{s}^{(k_3)},\bar{s}^{(k_4)})_{z_2}
\Bigg].
\end{eqnarray}
The normalization factor $\Lambda_c$ determines the conditional entropy, see Eq.~\eqref{HCC}. It has the following form:
\begin{eqnarray}\label{lambdac}
\Lambda_c=\left( \frac{T_0}{\pi Q L} \right)^{2M+1}\left[1
+\left(F^{k,k}_{2}-2G^{k,k'}_{1}H^{k',k}_{1}\right)\right] =\left(
\frac{T_0}{\pi Q L} \right)^{2M+1}\left[1+\gamma^2 L^2
{C}_{s_1}{C}_{s_2}\bar{{C}}_{s_3}\,\,\bar{{C}}_{s_4}J^{s_1,s_2;{s_3},{s_4}}_{\Lambda}\right],
\end{eqnarray}
where the real dimensionless coefficients
$J^{s_1,s_2;{s_3},{s_4}}_{\Lambda}$ are expressed in the following
way:
\begin{eqnarray}\label{Jlambda}
&&J^{s_1,s_2;{s_3},{s_4}}_{\Lambda} =\frac{1}{2 \pi^2
T^2_0}\int^{L}_0 \frac{dz_1}{L}\int^{L}_0 \frac{dz_2}{L}\Bigg[
\frac{z_1 z_2}{L^2}
(s^{(s_1)},s^{(s_2)};\bar{s}^{(r)},\bar{s}^{(r')})_{z_1}
(s^{(r')},s^{(r)};\bar{s}^{(s_3)},\bar{s}^{(s_4)})_{z_2} - \nonumber
\\&& {2\pi T_0} \frac{\min(z_1,z_2)}{ L} \int_{\tilde{W}} d \omega_a
(s^{(s_1)},s^{(s_2)};\omega_a,
\bar{s}^{(r)})_{z_1}(\omega_a,s^{(r)};\bar{s}^{(s_3)},\bar{s}^{(s_4)})_{z_2}
\Bigg].
\end{eqnarray}
Note, as noted previously, the sum over indexes $r$ and $r'$ is assumed here.

For the further numerical calculations can present this quantity as
follows
\begin{eqnarray}\label{Jlambda1}
J^{s_1,s_2;{s_3},{s_4}}_{\Lambda} = 2 b^{s_1,s_2;{r},{r'}}_1
b^{r',r;{s_3},{s_4}}_1-2 b^{s_1,s_2;{s_3},{s_4}}_2,
\end{eqnarray}
where
\begin{eqnarray}\label{b1koeff}
b^{s_1,s_2;{s_3},{s_4}}_1={\frac{1}{2\pi T_0}}\int^L_ 0 \frac{dz}{L}
 (s^{(s_1)},s^{(s_2)};\bar{s}^{(s_3)},\bar{s}^{(s_4)})_z \frac{z}{L},
\end{eqnarray}
\begin{eqnarray}\label{b2koeff}
b^{s_1,s_2;{s_3},{s_4}}_2={\frac{1}{2\pi T_0}} \int^{L}_0
\frac{dz_1}{L}\int^{L}_0 \frac{dz_2}{L} \frac{\min(z_1,z_2)}{ L}
\int_{\tilde{W}} d \omega_a
(\omega_a,s^{(r)};\bar{s}^{(s_3)},\bar{s}^{(s_4)})_{z_1}
(s^{(s_1)},s^{(s_2)};\omega_a, \bar{s}^{(r)})_{z_2}.
\end{eqnarray}
The method of the numerical calculation of these coefficients
$b^{s_1,s_2;{s_3},{s_4}}_1$ and $b^{s_1,s_2;{s_3},{s_4}}_2$ is
presented in the Appendix \ref{NumCalcSection}.

\section{Calculation of the coefficients $J_{I}^{s_1,s_2;{s_3},{s_4}}=J^{s_1,s_2;{s_3},{s_4}}+J_{\Lambda}^{s_1,s_2;{s_3},{s_4}}$}
\label{NumCalcSection}

In the following we will use the dimensionless dispersion parameter
\eqref{tildebeta}:
\begin{eqnarray}
\tilde{\beta}=\beta L W^2/2.
\end{eqnarray}

We choose the envelope of the signal in the sinc form
\eqref{Sincft}. Thus, the input signal $X(t)$, see
Eq.~(\ref{Xtmodelg}), has the form:
\begin{eqnarray}\label{sicXt}
X(t)=\sum_{k=-M}^M C_k \,\mathrm{sinc}\left[W(t-k T_0)/2\right],
\end{eqnarray}
where for the first model we use $T_0=2\pi/W$, here $W$ is given
frequency bandwidth of the input signal. The Fourier transform of
the signal (\ref{sicXt}) has the form
\begin{eqnarray}
X(\omega)\equiv \int^{\infty}_{-\infty}dt X(t) e^{i \omega t}
=\frac{2\pi}{W}\theta(W/2-|\omega|)\sum_{k=-M}^MC_k e^{ik\omega
T_0}.
\end{eqnarray}
 To obtain the coefficient $C_k$ we can use the following expressions
\begin{eqnarray}
C_k&=&\int\frac{dt}{T_0}X(t)\mathrm{sinc}\left[\frac{W}{2}(t-k T_0)\right],\\
C_k&=&\int^{W/2}_{-W/2}\frac{d\omega}{2\pi}X(\omega)e^{-i\omega k
T_0}.
\end{eqnarray}
Note, that the property (\ref{nonoverlapping}) is exact as resulting
from the orthogonality for the complete set of functions $s(t-k T_0)$:
\begin{eqnarray}\label{orthogonality}
\int^{\infty}_{-\infty} \frac{dt}{T_0} \mathrm{sinc}\left[W(t-k_1
T_0)/2\right] \mathrm{sinc}\left[W(t-k_2
T_0)/2\right]=\delta_{k_1,k_2}.
\end{eqnarray}

\subsection{Calculation of the Jacobian determinant via coefficients $J^{s_1,s_2;{s_3},{s_4}}$}

To find coefficients $J^{s_1,s_2;{s_3},{s_4}}$ from representation
(\ref{JJacobian4}):
\begin{eqnarray}
&&J^{s_1,s_2;{s_3},{s_4}}=2 a_1^{r',s_1;s_3,r}a_1^{r,s_2;s_4,r'}+2
a_1^{r',s_2;s_3,r}a_1^{r,s_1;s_4,r'}-a_1^{s_1,s_2;r,r'}a_1^{r,r';s_3,s_4}+\nonumber
\\&&\Big[A_2^{s_1,s_2,r;\,r,s_3,s_4}-A_2^{r,s_1,s_2;\,s_3,s_4,r}-A_2^{r,s_2,s_1;\,s_3,s_4,r}-
A_2^{r,s_1,s_2;\,s_4,s_3,r}-A_2^{r,s_2,s_1;\,s_4,s_3,r}\Big]+\overline{\Big[s_1
\leftrightarrow s_3, s_2 \leftrightarrow s_4\Big]}.
\end{eqnarray}
we should calculate two sums (here we write them explicitly):
$\sum_{r=-M}^M A_2^{r,s_1,s_2;\,{s_3},{s_4},{r}}$ and $\sum_{r=-M}^M
A_2^{s_1,s_2,r;\,{r},{s_3},{s_4}}$. This summation can be performed
at first under the integral, see Eq.~(\ref{A2form4}) below. The second sum
is automatically symmetric under substitutions $s_1 \leftrightarrow
s_2$, $s_3 \leftrightarrow s_4$.

To find the mutual information (\ref{IPopt}) one should find the sum
$J^{r,s;{r},{s}}+J^{r,s;{s},{r}}$:
\begin{eqnarray}\label{sumj}
&&J^{r,s;{r},{s}}+J^{r,s;{s},{r}}=2J^{r,s;{r},{s}}=\nonumber
\\&& 2 \mathrm{Re} \Big[2
a_1^{r_1,r;r,r_2}\overline{a_1^{r_1,r';r',r_2}}+a_1^{r_1,r_2;r_3,r_4}\overline{a_1^{r_1,r_2;r_3,r_4}}-
4A_2^{r,r',s;\,r',s,{r}}-2 A_2^{r,r',s;\,s,r',r}\Big].
\end{eqnarray}

\subsubsection{The calculation of the coefficients $a_1^{n,m;p,k}$}
For the sinc envelope \eqref{Sincft} the coefficient
$a_1^{n,m;p,k}$, see the definition in Eq.~(\ref{Akoeff}), can be
rewritten in the form:
\begin{eqnarray} \label{a1form1}
a_1^{n,m;p,k}=\frac{i}{8}\int_{-1}^1dxdx_1
dx_2\theta(1-|x_1+x_2-x|)e^{i\pi(x_1(n-p)+x_2(m-p)-x(k-p))}{\cal
G}(\tilde{\beta} (x_1-x)(x_2-x)),
\end{eqnarray}
where we have introduced the  function ${\cal G}(x)$:
\begin{eqnarray}\label{Gfunction}
{\cal G}(x)=-i \int^1_0 dz  e^{-i z
x}=\frac{\cos\left(x\right)-1}{x}-i\frac{\sin(x)-x}{x}-i=-i\sum^{\infty}_{k=0}
\frac{(-i x)^k}{k! (k+1)}.
\end{eqnarray}

It is obvious that for the real argument
\begin{eqnarray}\label{Gproperty}
 {\cal G}(x)=-\bar{{\cal G}}(-x).
\end{eqnarray}
Integration by part in (\ref{a1form1}) and the relation
(\ref{Gproperty}) allows us to reduce the representation
(\ref{a1form1}) to the following one
\begin{eqnarray}\label{a1form2}
&&a_1^{n,m;p,k}=\frac{i (-1)^{n+m-p-k}}{2\pi (n+m-p-k)}\int^2_0 dy
\int^{2-y}_0 dt\Big\{   {\cal G}(\tilde{\beta}  y
t)\cos\left[\frac{\pi}{2}(k-p)(t+y)\right]\sin\left[\frac{\pi
t}{2}(k+p-2m)+\frac{\pi y}{2}(k+p-2n)\right] +\nonumber \\&& \bar{{\cal
G}}(\tilde{\beta}  y
t)\cos\left[\frac{\pi}{2}(m-n)(t+y)\right]\sin\left[\frac{\pi
t}{2}(m+n-2p)+\frac{\pi y}{2}(m+n-2k)\right] \Big\},  \qquad \qquad
n+m-p-k \neq 0,
\end{eqnarray}
and if $n+m-k-p=0$ one has
\begin{eqnarray}\label{a1form2zero}
&&a_1^{n,m;p,k}=\frac{i}{2}\int^2_0 dy \int^{2-y}_0 dt (1-y)\Big\{
{\cal G}(\tilde{\beta}  y t)
\cos\left[\frac{\pi}{2}(k-p)(t+y)\right]\cos\left[\frac{\pi
t}{2}(k+p-2m)+\frac{\pi y}{2}(k+p-2n)\right]-\nonumber
\\&&\bar{{\cal G}}(\tilde{\beta}  y t)
\cos\left[\frac{\pi}{2}(m-n)(t+y)\right]\cos\left[\frac{\pi
t}{2}(m+n-2p)+\frac{\pi y}{2}(m+n-2k)\right]\Big\},
  \qquad \qquad n+m-p-k = 0.
\end{eqnarray}

\subsubsection{The calculation of the coefficients $A^{m_1,m_2,m_3;\,{m_4},{m_5},{m_6}}_2 $}

Quantities  $A^{m_1,m_2,m_3;\,{m_4},{m_5},{m_6}}_2 $ are presented through the
nine-fold integrals \eqref{A2form01}. For the sinc envelope \eqref{Sincft} they take the following form:
\begin{eqnarray}\label{A2form1}
&&A^{m_1,m_2,m_3;\,{m_4},{m_5},{m_6}}_2=\frac{1}{2\pi T_0}\int^L_ 0
\frac{dz_1}{L} \int^{z_1}_ 0 \frac{dz_2}{L} \int_{\tilde{W}}
d\omega_a
(\omega_a,s^{(m_3)};\bar{s}^{(m_5)},\bar{s}^{(m_6)})_{z_1}(s^{(m_1)},s^{(m_2)};\omega_a,\bar{s}^{(m_4)})_{z_2}=\nonumber
\\&& \int^{1}_{0} d \zeta_1 \int^{\zeta_1}_0 d \zeta_2
\int^{\infty}_{-\infty} dx  \int^{1/2}_{-1/2} d x_1 \ldots
\int^{1/2}_{-1/2}d x_6 \delta(x_1+x_2-x-x_4) \delta(x+x_3-x_5-x_6)
\times \nonumber \\&& \exp\left[2 i \tilde{\beta} \zeta_1
(x^2+x^2_3-x^2_5-x^2_6)+2 i \tilde{\beta} \zeta_2
(x^2_1+x^2_2-x^2-x^2_4) \right] e^{2\pi i (m_1 x_1 +m_2 x_2 +m_3 x_3
-m_4 x_4-m_5 x_5 -m_6 x_6)}.
\end{eqnarray}
Here we have used that the noise bandwidth $\tilde{W} \gg W$  and
introduced the dimensionless variables $\zeta_{1,2}=z_{1,2}/L$,
$x_i=\omega_i/W$, $x=\omega_a/W$.

For the case of large $\tilde{\beta} $ it is convenient to present
(\ref{A2form1}) in the form where both delta functions in
Eq.~(\ref{A2form1}) are integrated out and the the quadratic form is
reduced to a diagonal form. To this end, we set $x_6$ equal to
$x_1+x_2+x_3-x_4-x_5$ and perform the change of variables:
\begin{eqnarray}
x_1=y_5,\quad x_2=2y_4+y_5, \quad x_3=2 y_2+y_3+y_4+y_5, \quad
x_4=-y_3+y_4+y_5, \quad x_5=y_1+y_2+y_3+y_4+y_5.
\end{eqnarray}
These transformations of Eq.~(\ref{A2form1}) lead  {to the
first representation} of the quantity (\ref{A2form1})
\begin{eqnarray}\label{A2form3s}
&&A^{m_1,m_2,m_3;\,{m_4},{m_5},{m_6}}_2= 4 \int^{1}_{0} d \zeta_1
\int^{\zeta_1}_0 d \zeta_2 \int^{\frac{1}{2}}_{-\frac{1}{2}} d y_5
\int^{\frac{1}{4}-\frac{y_5}{2}}_{-\frac{1}{4}-\frac{y_5}{2}} d y_4
\int^{\frac{1}{2}+y_4+y_5}_{-\frac{1}{2}+y_4+y_5} \!\!\!\!\!\!d y_3
\int^{\min\{\frac{1}{4}-
\frac{y_3+y_4+y_5}{2},\frac{1}{2}-y_3-y_4-y_5\}}_{\max[-\frac{1}{4}-
\frac{y_3+y_4+y_5}{2},-\frac{1}{2}-y_3-y_4-y_5]}
\!\!\!\!\!\!\!\!\!\!\!\!\!\!\!\!\!\!\!\!\!\!\!\!d y_2 \times
\nonumber \\&&
\int^{\frac{1}{2}-|y_2+y_3+y_4+y_5|}_{-\frac{1}{2}+|y_2+y_3+y_4+y_5|}
d y_1  \exp\left[4 i \tilde{\beta} \zeta_1 (-y^2_1+y^2_2)+ 4 i
\tilde{\beta} \zeta_2 (-y^2_3+y^2_4)\right] e^{2\pi i y_5
(m_1+m_2+m_3-m_4-m_5-m_6)}\times \nonumber \\&& e^{-2\pi i y_1
(m_5-m_6)+2\pi i y_2 (2m_3-m_5-m_6)+2 \pi i y_3
(m_3+m_4-m_5-m_6)+2\pi i y_4 (2m_2+m_3-m_4-m_5-m_6)}.
\end{eqnarray}
Note that for inner integral (over $dy_i$) the point
$(y_1=0,y_2=0,y_3=0,y_4=0, y_5)$ is  always inside the integration
domain.
We can perform integration in Eq.~(\ref{A2form3s}) over $\zeta_1$
and $\zeta_2$,  and then integrate by part over $y_5$. Thus, this
procedure results in nine four-fold integrals:
\begin{eqnarray}\label{A2sum}
A^{m_1,m_2,m_3;\,{m_4},{m_5},{m_6}}_2=\sum_{i=1}^{9}I^{\{m\}}_i.
\end{eqnarray}
We introduce $N=m_1+m_2+m_3-m_4-m_5-m_6$. For the case $N \neq 0$
the terms $I^{\{m\}}_i$ have the following form:
\begin{eqnarray}\label{Im1}
I^{\{m\}}_1&=&\frac{2e^{i\pi N}}{i\pi
N}\int\limits^{0}_{-1/2}dy_4\int\limits^{1+y_4}_{y_4}dy_3\int\limits^{\min\{-(y_3+y_4)/2,-(y_3+y_4)\}}_{\max\{-1/2-(y_3+y_4)/2,-1-(y_3+y_4)\}}dy_2
\int\limits^{1/2-|y_2+y_3+y_4+1/2|}_{-1/2+|y_2+y_3+y_4+1/2|}dy_1\times\nonumber\\&&{\cal
G}_2(y_2^2-y_1^2,y_4^2-y_3^2)Ex^{\{m\}}(y_4,y_3,y_2,y_1),
\end{eqnarray}
\begin{eqnarray}%
I^{\{m\}}_2&=&-\frac{2e^{-i\pi N}}{i\pi N}\int\limits^{1/2}_{0}dy_4\int\limits^{y_4}_{-1+y_4}dy_3\int\limits^{\min\{1/2-(y_3+y_4)/2, 1-(y_3+y_4)\}}_{\max\{-(y_3+y_4)/2,-(y_3+y_4)\}}dy_2\int\limits^{1/2-|y_2+y_3+y_4-1/2|}_{-1/2+|y_2+y_3+y_4-1/2|}dy_1 \times\nonumber\\
&&{\cal G}_2(y_2^2-y_1^2,y_4^2-y_3^2)Ex^{\{m\}}(y_4,y_3,y_2,y_1),
\end{eqnarray}
\begin{eqnarray}\label{Im3}%
I^{\{m\}}_3&=&\frac{1}{i\pi N}\int\limits^{1/2}_{-1/2}e^{2i\pi Ny_5}dy_5\int\limits^{\frac{3}{4}+\frac{y_5}{2}}_{-\frac{1}{4}+\frac{y_5}{2}}dy_3\int\limits^{\min\{\frac{1}{8}-\frac{2y_3+y_5}{4},\frac{1}{4}-\frac{2y_3+y_5}{2}\}}_{\max\{-\frac{3}{8}-\frac{2y_3+y_5}{4},-\frac{3}{4}-\frac{2y_3+y_5}{2}\}}dy_2 \int\limits^{1/2-|y_2+y_3+\frac{y_5}{2}+\frac{1}{4}|}_{-1/2+|y_2+y_3+\frac{y_5}{2}+\frac{1}{4}|}dy_1 \times\nonumber\\
&& {\cal
G}_2(y_2^2-y_1^2,(1/4-y_5/2)^2-y_3^2)Ex^{\{m\}}\left(\frac{1}{4}-\frac{y_5}{2},y_3,y_2,y_1\right),
\end{eqnarray}
\begin{eqnarray}%
I^{\{m\}}_4&=&-\frac{1}{i\pi N}\int\limits^{1/2}_{-1/2}e^{2i\pi Ny_5}dy_5\int\limits^{\frac{1}{4}+\frac{y_5}{2}}_{-\frac{3}{4}+\frac{y_5}{2}}dy_3\int\limits^{\min\{\frac{3}{8}- \frac{2y_3+y_5}{4},\frac{3}{4}-\frac{2y_3+y_5}{2}\}}_{\max\{-\frac{1}{8}-\frac{2y_3+y_5}{4},-\frac{1}{4}-\frac{2y_3+y_5}{2}\}} dy_2\int\limits^{1/2-|y_2+y_3+\frac{y_5}{2}-\frac{1}{4}|}_{-1/2+|y_2+y_3+\frac{y_5}{2}-\frac{1}{4}|}dy_1 \times\nonumber\\
&& {\cal
G}_2(y_2^2-y_1^2,(1/4+y_5/2)^2-y_3^2)Ex^{\{m\}}\left(-\frac{1}{4}-\frac{y_5}{2},y_3,y_2,y_1\right),
\end{eqnarray}
\begin{eqnarray}%
I^{\{m\}}_5&=&-\frac{2}{i\pi N}\int\limits^{1/2}_{-1/2}e^{2i\pi Ny_5}dy_5\int\limits^{\frac{1}{4}-\frac{y_5}{2}}_{-\frac{1}{4}-\frac{y_5}{2}}dy_4\int \limits^{\min\{-y_4-y_5,-2(y_4+y_5)\}}_{\max\{-1/2-y_4-y_5,-1-2(y_4+y_5)\}}dy_2\int\limits^{1/2-|y_2+2y_4+2y_5+1/2|}_{-1/2+|y_2+2y_4+2y_5+1/2|}dy_1 \times\nonumber\\
&& {\cal
G}_2(y_2^2-y_1^2,y_4^2-(y_4+y_5+1/2)^2)Ex^{\{m\}}\left(y_4,\frac{1}{2}+y_4+y_5,y_2,y_1\right),
\end{eqnarray}
\begin{eqnarray}%
I^{\{m\}}_6&=&\frac{2}{i\pi N}\int\limits^{1/2}_{-1/2}e^{2i\pi Ny_5}dy_5\int\limits^{\frac{1}{4}-\frac{y_5}{2}}_{-\frac{1}{4}-\frac{y_5}{2}}dy_4 \int\limits^{\min\{1/2-y_4-y_5,1-2(y_4+y_5)\}}_{\max\{-y_4-y_5,-2(y_4+y_5)\}}dy_2\int\limits^{1/2-|y_2+2y_4+2y_5-1/2|}_{-1/2+|y_2+2y_4+2y_5-1/2|}dy_1 \times\nonumber\\
&&{\cal
G}_2(y_2^2-y_1^2,y_4^2-(y_4+y_5-1/2)^2)Ex^{\{m\}}\left(y_4,-\frac{1}{2}+y_4+y_5,y_2,y_1\right),
\end{eqnarray}
\begin{eqnarray}%
I_7&=&\frac{1}{i\pi N}\int\limits^{1/2}_{-1/2}e^{2i\pi Ny_5}dy_5\int\limits^{\frac{1}{4}-\frac{y_5}{2}}_{-\frac{1}{4}-\frac{y_5}{2}}dy_4\int \limits^{\frac{1}{2}+y_4+y_5}_{-\frac{1}{2}+y_4+y_5}dy_3\,\theta\left(\frac{1}{2}-(y_3+y_4+y_5)\right)\int\limits^{\frac{1}{2}-|\frac{1}{4}+\frac{y_3+y_4+y_5}{2}|}_{-\frac{1}{2}+|\frac{1}{4}+\frac{y_3+y_4+y_5}{2}|}dy_1 \times\nonumber\\
&&{\cal
G}_2\left(\left(\frac{1}{4}-\frac{y_3+y_4+y_5}{2}\right)^2-y_1^2,y_4^2-y_3^2\right)Ex^{\{m\}}\left(y_4,y_3,\frac{1}{4}-\frac{y_3+y_4+y_5}{2},y_1\right),
\end{eqnarray}
\begin{eqnarray}
I^{\{m\}}_8&=&-\frac{1}{i\pi N}\int\limits^{1/2}_{-1/2}e^{2i\pi Ny_5}dy_5\int\limits^{\frac{1}{4}-\frac{y_5}{2}}_{-\frac{1}{4}-\frac{y_5}{2}}dy_4\int \limits^{\frac{1}{2}+y_4+y_5}_{-\frac{1}{2}+y_4+y_5}dy_3\,\theta\left(\frac{1}{2}+y_3+y_4+y_5\right)\int\limits^{\frac{1}{2}-|-\frac{1}{4}+\frac{y_3+y_4+y_5}{2}|}_{-\frac{1}{2}+|-\frac{1}{4}+\frac{y_3+y_4+y_5}{2}|}dy_1 \times\nonumber\\
&&{\cal
G}_2\left(\left(\frac{1}{4}+\frac{y_3+y_4+y_5}{2}\right)^2-y_1^2,y_4^2-y_3^2\right)Ex^{\{m\}}\left(y_4,y_3,-\frac{1}{4}-\frac{y_3+y_4+y_5}{2},y_1\right),
\end{eqnarray}
\begin{eqnarray} \label{Im9} 
I^{\{m\}}_{9}&=&\frac{2}{i\pi N}\int\limits^{1/2}_{-1/2}e^{2i\pi Ny_5}dy_5\int\limits^{\frac{1}{4}-\frac{y_5}{2}}_{-\frac{1}{4}-\frac{y_5}{2}}dy_4 \int\limits^{\frac{1}{2}+y_4+y_5}_{-\frac{1}{2}+y_4+y_5}dy_3\int\limits^{\min\{\frac{1}{4}-\frac{y_3+y_4+y_5}{2},\frac{1}{2}-(y_3+y_4+y_5)\}}_{\max\{-\frac{1}{4}-\frac{y_3+y_4+y_5}{2},-\frac{1}{2}-(y_3+y_4+y_5)\}}dy_2 \mathrm{sign}(y_2+y_3+y_4+y_5)\times\nonumber\\
&&{\cal G}_2\left(y_2^2-\left(\frac{1}{2}-|y_2+y_3+y_4+y_5|\right)^2,y_4^2-y_3^2\right)\Bigg[Ex^{\{m\}}\left(y_4,y_3,y_2,\frac{1}{2}-|y_2+y_3+y_4+y_5|\right)+\nonumber\\
&&Ex^{\{m\}}\left(y_4,y_3,y_2,-\frac{1}{2}+|y_2+y_3+y_4+y_5|\right)\Bigg],
\end{eqnarray}
where we have introduced the functions
\begin{eqnarray}
Ex^{\{m\}}(y_4,y_3,y_2,y_1)=e^{-2\pi i y_1 (m_5-m_6)+2\pi i y_2
(2m_3-m_5-m_6)+2 \pi i y_3 (m_3+m_4-m_5-m_6)+2\pi i y_4
(2m_2+m_3-m_4-m_5-m_6)},
\end{eqnarray}
\begin{eqnarray}\label{G2ab}
{\cal G}_2(a,b)=\int^1_0 d\zeta_1 \int^{\zeta_1}_0 d\zeta_2
 e^{4i \tilde{\beta} (\zeta_1 a+ \zeta_2
b)}=-\frac{1}{16\tilde{\beta}^2b}\left(\frac{e^{4i\tilde{\beta}(a+b)}-1}{a+b}-\frac{e^{4i\tilde{\beta}a}-1}{a}\right).
\end{eqnarray}

For the case $N=0$ we obtain:
\begin{eqnarray} \label{Im1mod}%
I^{\{m\}}_1&=&2\int\limits^{0}_{-1/2}dy_4\int\limits^{1+y_4}_{y_4}dy_3\int \limits^{\min\{-(y_3+y_4)/2,-(y_3+y_4)\}}_{\max\{-1/2-(y_3+y_4)/2,-1-(y_3+y_4)\}}dy_2\int\limits^{1/2-|y_2+y_3+y_4+1/2|}_{-1/2+|y_2+y_3+y_4+1/2|}dy_1 \times\nonumber\\
&&{\cal G}_2(y_2^2-y_1^2,y_4^2-y_3^2)Ex^{\{m\}}(y_4,y_3,y_2,y_1),
\end{eqnarray}
\begin{eqnarray} \label{Im2mod}
I^{\{m\}}_2&=&2\int\limits^{1/2}_{0}dy_4\int\limits^{y_4}_{-1+y_4}dy_3 \int\limits^{\min\{1/2-(y_3+y_4)/2,1-(y_3+y_4)\}}_{\max\{-(y_3+y_4)/2,-(y_3+y_4)\}}dy_2\int\limits^{1/2-|y_2+y_3+y_4-1/2|}_{-1/2+|y_2+y_3+y_4-1/2|}dy_1 \times\nonumber\\
&&{\cal G}_2(y_2^2-y_1^2,y_4^2-y_3^2)Ex^{\{m\}}(y_4,y_3,y_2,y_1),
\end{eqnarray}
and other integrals $I_{3}-I_{11}$ for the case $N=0$ can be
obtained from ones in  the case $N\neq 0$ by the change
$\dfrac{e^{2i\pi N y_5}}{2\pi i N} \rightarrow  y_5$ under the
integral over $y_5$.

{The second representation} of the quantity (\ref{A2form1})
reads as the four-fold integral resulting from the eliminating
$\delta$-functions in Eq.~(\ref{A2form1}) by virtue of the integral
representation (specifically,
$\delta(x+x_3-x_5-x_6)=\int^{\infty}_{-\infty} d\alpha_1 \exp[2\pi i
\alpha_1 (x+x_3-x_5-x_6)]$) followed by  the Gaussian integration
over  $x$, and the integration (see, Eq.~(\ref{Eabfunction}) below)
over $x_i$:
\begin{eqnarray}\label{A2form3sec}
&&A^{m_1,m_2,m_3;\,{m_4},{m_5},{m_6}}_2= \int^{1}_{0} d \zeta_1
\int^{\zeta_1}_0 d \zeta_2 \int^{\infty}_{-\infty} d\alpha_1
\int^{\infty}_{-\infty} d\alpha_2 \sqrt{\frac{\pi}{2 \tilde{\beta}
(\zeta_1-\zeta_2) }}\exp\left[-i \frac{\pi^2
(\alpha_1-\alpha_2)^2}{2\tilde{\beta} (\zeta_1-\zeta_2)}+i
\frac{\pi}{4}\right]\times \nonumber \\&&
E(2\pi(\alpha_1+m_3),2\tilde{\beta}\zeta_1)\tilde{E}(2\pi(\alpha_1+m_5),2\tilde{\beta}\zeta_1)
\tilde{E}(2\pi(\alpha_1+m_6),2\tilde{\beta}\zeta_1) \times \nonumber
\\&&
E(2\pi(\alpha_2+m_1),2\tilde{\beta}\zeta_2)E(2\pi(\alpha_2+m_2),2\tilde{\beta}\zeta_2)
\tilde{E}(2\pi(\alpha_2+m_4),2\tilde{\beta}\zeta_2),
\end{eqnarray}
where we have introduced the functions
\begin{eqnarray}
E(a,b)=\int^{1/2}_{-1/2} d y
e^{i b y^2+ i a y}, \qquad \qquad \tilde{E}(a,b)=\int^{1/2}_{-1/2} d y e^{-i b
y^2-i a y},
\end{eqnarray}
where one has the following representations for
$E(a,b)$ in the case of positive $b>0$:
\begin{eqnarray}\label{Eabfunction}
E(a,b)=\int^{\sqrt{b}/2}_{-\sqrt{b}/2} \!\!\!\frac{du}{\sqrt{b}} \, e^{i  u^2+ i \frac{a}{\sqrt{b}} u}=e^{-i b \frac{\partial^2}{\partial a^2}}\mathrm{sinc}\left(\frac{a}{2}\right)&=-\sqrt{\dfrac{\pi}{b}}\exp\left[-i \dfrac{a^2}{4 b}+ i \dfrac{\pi}{4}\right] \dfrac{\mathrm{erf}\left(e^{i \frac{3\pi}{4}}\frac{b-a}{2\sqrt{b}}\right)+\mathrm{erf}\left(e^{i \frac{3\pi}{4}}\frac{b+a}{2\sqrt{b}}\right)}{2}, 
\end{eqnarray}
here $\mathrm{erf}(x)=\frac{2}{\sqrt{\pi}}\int^{x}_0 dt e^{-t^2}$ is
standard Gauss error function. It is worth noting that the
convergence of the improper integral (\ref{A2form3sec}) is provided by
the following asymptotical behavior of the function $E(a,b)$ for the
large $a$:
\begin{eqnarray}\label{Eabfunctionaslargea}
E(a,b)\sim  \,e^{i {b}/{4}}\mathrm{sinc}(a/2)+{\cal O}(1/a^2),
\end{eqnarray}
which directly follows from the  second representation in
Eq.~(\ref{Eabfunction}). Note that
$\tilde{E}(a,b)=\overline{E(a,b)}$ only for the real arguments.


There are some difficulties in the numerical calculations of the
error function of a complex variable. To avoid these difficulties we
use the method of the approximate calculation described in
Ref.~\cite{Salzer:1951}. From the representation (3) in
Ref.~\cite{Salzer:1951} we have the approximate identity
\begin{eqnarray}\label{expapprox}
e^{i u^2}\approx
\frac{1}{\sqrt{\pi}}\left(\frac{1}{2}+\sum^{\infty}_{n=1}e^{-n^2/4}\cosh(n
u e^{i \pi/4})\right),
\end{eqnarray}
where the accuracy of the approximation is on the level $10^{-7}$
for all complex $u$: $|u|<2.9$. Using the first representation in
Eq.~(\ref{Eabfunction}) one arrives at approximation
\begin{eqnarray}\label{Eabapprox}
\!\!\!E(a,b) \approx
\frac{1}{\sqrt{\pi}}\left(\frac{\mathrm{sinc}(\frac{a}{2})}{2}+2\,
\sum^{\infty}_{n=1}\frac{a \sin \left(\frac{a}{2}\right)
\cosh\left(\frac{\sqrt{b}}{2}  e^{i \pi/4}n\right)+\sqrt{b}\,
\cos\left( \frac{a}{2}\right) e^{i \pi/4} n
\sinh\left(\frac{\sqrt{b}}{2}  e^{i \pi/4}n\right)}{a^2+ i b\,
n^2}\,e^{-\dfrac{n^2}{4}}\right),
\end{eqnarray}
where for $|a|<100$ and $|b|<15$ the accuracy of the approximation
(\ref{Eabapprox}) is on the level $10^{-7}$.

To avoid the oscillating function when numerically integrating  we
perform the obvious change of variables and the rotation in the
complex plane in Eq.~(\ref{A2form3sec}). In this wise we arrive at the
following representation of the integral (\ref{A2form3sec}):
\begin{eqnarray}\label{A2form4}
&&A^{m_1,m_2,m_3;\,{m_4},{m_5},{m_6}}_2=\int^{1}_{0} d t_1\, t_1
\int^{1}_0 d t_2 \int^{\infty}_{-\infty}  \frac{d \alpha}{2 \pi}
\int^{\infty}_{-\infty} d  \zeta \frac{e^{-\zeta^2}}{\sqrt{\pi}}
E\left(2\pi m_1+\alpha-\zeta \sqrt{2 t_1 t_2
\tilde{\beta}}e^{-i\frac{\pi}{4}},2\tilde{\beta}t_1(1-t_2)\right)\times
\nonumber \\&& E\left(2\pi m_2+\alpha-\zeta \sqrt{2 t_1 t_2
\tilde{\beta}}e^{-i\frac{\pi}{4}},2\tilde{\beta}t_1(1-t_2)\right)
E\left(2\pi m_3+\alpha+\zeta \sqrt{2 t_1 t_2
\tilde{\beta}}e^{-i\frac{\pi}{4}},2\tilde{\beta}t_1\right) \times
\nonumber \\&& \tilde{E}\left(2\pi m_4+\alpha-\zeta \sqrt{2 t_1 t_2
\tilde{\beta}}e^{-i\frac{\pi}{4}},2\tilde{\beta}t_1(1-t_2)\right)
\tilde{E}\left(2\pi m_5+\alpha+\zeta \sqrt{2 t_1 t_2
\tilde{\beta}}e^{-i\frac{\pi}{4}},2\tilde{\beta}t_1\right)\times
\nonumber \\&& \tilde{E}\left(2\pi m_6+\alpha+\zeta \sqrt{2 t_1 t_2
\tilde{\beta}}e^{-i\frac{\pi}{4}},2\tilde{\beta}t_1\right),
\end{eqnarray}
where one can use the approximation (\ref{Eabapprox}) for the
dispersion parameter $\tilde{\beta}< 7.5$ (with the guaranteed
approximation level $10^{-7}$).

\subsection{Calculation  of the
normalization factor via coefficients
$J^{s_1,s_2;{s_3},{s_4}}_{\Lambda}$}

Here we present the calculation strategy for the contribution of the
normalization factor (\ref{lambdac}) to the mutual information, i.e.
the quantities $J^{s_1,s_2;{s_3},{s_4}}_{\Lambda}$, see
Eq.~(\ref{Jlambda1}):
$$
J^{s_1,s_2;{s_3},{s_4}}_{\Lambda} = 2 b^{s_1,s_2;{r},{r'}}_1
b^{r',r;{s_3},{s_4}}_1-2 b^{s_1,s_2;{s_3},{s_4}}_2,
$$
where coefficients $b^{n,m;p,k}_1$ and $b^{k_1,k_2;k_3,k_4}_2$ are
defined in the Eqs. \eqref{b1koeff}, \eqref{b2koeff} respectively.

To find the mutual information (\ref{IPopt}) one should find the
following sum
\begin{eqnarray}\label{sumjlambda}
J^{r,s;{r},{s}}_{\Lambda}+J^{r,s;{s},{r}}_{\Lambda}=2J^{r,s;{r},{s}}_{\Lambda}=
 4 b^{m,p;{r},{r'}}_1
b^{r',r;{m},{p}}_1-4 b^{m,p;{m},{p}}_2.
\end{eqnarray}
We perform our calculation for the sin envelope \eqref{Sincft}.

\subsubsection{The calculation of the coefficients $b_1^{n,m;p,k}$}
The method of the calculation of the coefficients (\ref{b1koeff})
is identical with the method of calculation of the coefficients
$a_1^{n,m;p,k}$. Thus, we introduce the auxilary function
\begin{eqnarray}\label{G1function}
{\cal G}_1(x)=-i \int^1_0 dz z e^{-i z x}=\frac{\cos x
-1}{x}+\frac{x-\sin x}{x^2}+i\frac{1-\cos x -x \sin
x}{x^2}=-i\sum^{\infty}_{k=0} \frac{(-i x)^k}{k! (k+2)} ,
\end{eqnarray}
It is obvious that for the real argument
\begin{eqnarray}\label{G1property}
 {\cal G}_1(x)=-\bar{\cal G}_1(-x).
\end{eqnarray}
We have from the definition \eqref{b1koeff}:
\begin{eqnarray} \label{b1form1}
b_1^{n,m;p,k}=\frac{i}{8}\int_{-1}^1dxdx_1
dx_2\theta(1-|x_1+x_2-x|)e^{i\pi(x_1(n-p)+x_2(m-p)-x(k-p))}{\cal
G}_1(\tilde{\beta} (x_1-x)(x_2-x)).
\end{eqnarray}
Integration by part in (\ref{b1form1}) and the relation
(\ref{G1property}) allows us to reduce the representation
(\ref{b1form1}) to the following one
\begin{eqnarray}\label{b1form2}
&&b_1^{n,m;p,k}=\frac{i (-1)^{n+m-p-k}}{2\pi (n+m-p-k)}\int^2_0 dy
\int^{2-y}_0 dt\Big\{   {\cal G}_1(\tilde{\beta}  y
t)\cos\left[\frac{\pi}{2}(k-p)(t+y)\right]\sin\left[\frac{\pi
t}{2}(k+p-2m)+\frac{\pi y}{2}(k+p-2n)\right] +\nonumber \\&& \bar{\cal
G}_1(\tilde{\beta}  y
t)\cos\left[\frac{\pi}{2}(m-n)(t+y)\right]\sin\left[\frac{\pi
t}{2}(m+n-2p)+\frac{\pi y}{2}(m+n-2k)\right] \Big\},  \qquad \qquad
n+m-p-k \neq 0,
\end{eqnarray}
and if $n+m-k-p=0$ one has
\begin{eqnarray}\label{b1form2zero}
&&b_1^{n,m;p,k}=\frac{i}{2}\int^2_0 dy \int^{2-y}_0 dt (1-y)\Big\{
{\cal G}_1(\tilde{\beta}  y t)
\cos\left[\frac{\pi}{2}(k-p)(t+y)\right]\cos\left[\frac{\pi
t}{2}(k+p-2m)+\frac{\pi y}{2}(k+p-2n)\right]-\nonumber
\\&&\bar{\cal G}_1(\tilde{\beta}  y t)
\cos\left[\frac{\pi}{2}(m-n)(t+y)\right]\cos\left[\frac{\pi
t}{2}(m+n-2p)+\frac{\pi y}{2}(m+n-2k)\right]\Big\},
  \qquad \qquad n+m-p-k = 0.
\end{eqnarray}
These two-fold integrals \eqref{b1form2} \eqref{b1form2zero} and  are easy to numerically calculate by the standard methods.

\subsubsection{The calculation of the coefficients $b_2^{k_1,k_2;k_3,k_4}$}
Now we proceed to the numerical calculation of the coefficients
$b_2^{k_1,k_2;k_3,k_4}$ defined in Eq.~\eqref{b2koeff}:
$$
b^{k_1,k_2;{k_3},{k_4}}_2={\frac{1}{2\pi T_0}} \int^{L}_0
\frac{dz_1}{L}\int^{L}_0 \frac{dz_2}{L} \frac{\min(z_1,z_2)}{ L}
\int_{\tilde{W}} d \omega_a
(\omega_a,s^{(r)};\bar{s}^{(k_3)},\bar{s}^{(k_4)})_{z_1}
(s^{(k_1)},s^{(k_2)};\omega_a, \bar{s}^{(r)})_{z_2},
$$
where $\min(z_1,z_2)=z_1 \theta(z_2-z_1)+z_2 \theta(z_1-z_2)$. Note,
this representation for the coefficients $b^{k_1,k_2;\,k_3,k_4}_2$
differs from the representation (\ref{A2form1}) for the coefficients
$A^{m_1,m_2,m_3;\,{m_4},{m_5},{m_6}}_2$ by the change of the
integrations over $z_1$ and $z_2$ ($\int^L_ 0 \frac{dz_1}{L}
\int^{z_1}_ 0 \frac{dz_2}{L} \rightarrow \int^L_0 \frac{dz_1}{L}
\int^{L}_0 \frac{dz_2}{L} \frac{\min(z_1,z_2)}{L}$), and by the
changes of indexes  $m_1 \rightarrow k_1$, $m_2 \rightarrow k_2$,
$m_3 \rightarrow r$, $m_4 \rightarrow r$, $m_5 \rightarrow k_3$,
$m_6 \rightarrow k_4$ followed by the summation over $r$ from $-M$
to $M$. Therefore, we can obtain the following representation from
Eq.~\eqref{A2form1}:
\begin{eqnarray}\label{b2form1}
&&b^{k_1,k_2;\,k_3,k_4}_2=\frac{1}{2\pi T_0}\int^L_ 0 \frac{dz_1}{L}
\int^{1}_ 0 \frac{dz_2}{L} \frac{\min(z_1,z_2)}{L}\int_{\tilde{W}}
d\omega_a
(\omega_a,s^{(r)};\bar{s}^{(k_3)},\bar{s}^{(k_4)})_{z_1}(s^{(k_1)},s^{(k_2)};\omega_a,\bar{s}^{(r)})_{z_2}=\nonumber
\\&& \int^{1}_{0} d \zeta_1 \int^{1}_0 d \zeta_2 \frac{\min(\zeta_1,\zeta_2)}{L}
\int^{\infty}_{-\infty} dx  \int^{1/2}_{-1/2} d x_1 \ldots
\int^{1/2}_{-1/2}d x_6 \delta(x_1+x_2-x-x_4) \delta(x+x_3-x_5-x_6)
\times \nonumber \\&& \exp\left[2 i \tilde{\beta} \zeta_1
(x^2+x^2_3-x^2_5-x^2_6)+2 i \tilde{\beta} \zeta_2
(x^2_1+x^2_2-x^2-x^2_4) \right] \sum^{M}_{r=-M} e^{2\pi i (k_1 x_1
+k_2 x_2 +r (x_3 - x_4)-k_3 x_5 -k_4 x_6)}.
\end{eqnarray}
Here we have used that the noise bandwidth $\tilde{W} \gg W$  and
employed the dimensionless variables $\zeta_{1,2}=z_{1,2}/L$,
$x_i=\omega_i/W$, $x=\omega_a/W$.

The \textbf{first representation } of the coefficients
$b^{k_1,k_2;\,k_3,k_4}_2$ reads similar to Eqs.~\eqref{A2form3s} and
\eqref{A2sum}
\begin{eqnarray}
b^{k_1,k_2;\,k_3,k_4}_2=\sum_{i=1}^{9}\tilde{I}^{\{k\}}_i,
\end{eqnarray}
where four-fold integrals $\tilde{I}^{\{k\}}_i$ can be obtained from
Eqs. \eqref{Im1}-- \eqref{Im9} and \eqref{Im1mod}, \eqref{Im2mod} by
the changes
\begin{eqnarray}\label{change1}
N \rightarrow \tilde{N}=k_1+k_2-k_3-k_4,\qquad {\cal G}_2(a,b)
\rightarrow {\cal G}_3(a,b), \qquad Ex^{\{m\}}(y_4,y_3,y_2,y_1)
\rightarrow Ex_{\{k\}}(y_4,y_3,y_2,y_1),
\end{eqnarray}
where
\begin{eqnarray}
Ex_{\{k\}}(y_4,y_3,y_2,y_1)=\sum^{M}_{r=-M} e^{-2\pi i y_1
(k_3-k_4)+2\pi i (y_2+y_3) (2 r-k_3-k_4)+2\pi i y_4 (2k_2-k_3-k_4)},
\end{eqnarray}
and
\begin{eqnarray}
&&{\cal G}_3(a,b)=\int^1_0 d\zeta_1 \int^1_0 d\zeta_2
\min(\zeta_1,\zeta_2) e^{4i \tilde{\beta} (\zeta_1 a+ \zeta_2
b)}=\nonumber\\&& \frac{i}{64 \tilde{\beta}^3 a^2 b^2
(a+b)}\left[e^{4 i \tilde{\beta} (a+b)}\left(4i \tilde{\beta} a b
(a+b) -a^2-a b-b^2\right) +(a+b)(a e^{4 i \tilde{\beta} a}+b e^{4 i
\tilde{\beta} b})-a b\right].
\end{eqnarray}
For the case $\tilde{N}=k_1+k_2-k_3-k_4 \neq 0$ we use the
\eqref{change1} for Eqs. \eqref{Im1}--\eqref{Im9} to obtain
$\tilde{I}^{\{k\}}_i$.  For the case $\tilde{N}=0$ we use the change
\eqref{change1} for Eqs. \eqref{Im1mod} and \eqref{Im2mod} to obtain
$\tilde{I}^{\{k\}}_1$ and  $\tilde{I}^{\{k\}}_2$, respectively, and
the change $\dfrac{e^{2i\pi N y_5}}{2\pi i N} \rightarrow  y_5$
under the integral over $y_5$ together with the change
\eqref{change1} in Eqs. \eqref{Im3}--\eqref{Im9} to obtain others
$\tilde{I}^{\{k\}}_i$ ($3 \leq i\leq 9$).

\textbf{The second representation} of the quantity (\ref{b2form1})
reads as the four-fold integral:
\begin{eqnarray}\label{b2form3sec}
&&b^{k_1,k_2;\,k_3,k_4}_2= \int^{1}_{0} d \zeta_1 \int^{1}_0 d
\zeta_2 \min(\zeta_1,\zeta_2) \int^{\infty}_{-\infty} d\alpha_1
\int^{\infty}_{-\infty} d\alpha_2 \sqrt{\frac{\pi}{2 \tilde{\beta}
|\zeta_1-\zeta_2| }}\exp\left[-i \frac{\pi^2
(\alpha_1-\alpha_2)^2}{2\tilde{\beta} (\zeta_1-\zeta_2)}+i
\frac{\pi}{4}{\mathrm{sign}(\zeta_1-\zeta_2)}\right]\times \nonumber
\\&& \sum^{M}_{r=-M} E(2\pi(\alpha_1+r),2\tilde{\beta}\zeta_1)
\tilde{E}(2\pi(\alpha_2+r),2\tilde{\beta}\zeta_2) \times \nonumber\\&&
E(2\pi(\alpha_2+k_1),2\tilde{\beta}\zeta_2)E(2\pi(\alpha_2+k_2),2\tilde{\beta}\zeta_2)
\tilde{E}(2\pi(\alpha_1+k_3),2\tilde{\beta}\zeta_1)
\tilde{E}(2\pi(\alpha_1+k_4),2\tilde{\beta}\zeta_1),
\end{eqnarray}
where functions $E(a,b)=\int^{1/2}_{-1/2} d y e^{i b y^2+ i a y}$
and  $\tilde{E}(a,b)=\int^{1/2}_{-1/2} d y e^{-i b y^2-i a y}$ are
defined through Eq.~\eqref{Eabfunction}. Let us stress once
\begin{eqnarray}
\tilde{E}(a,b)=\overline{E(\bar{a},\bar{b})},
\end{eqnarray}
where the overline means the complex conjugation, i.e., $\tilde{E}(a,b)=
\overline{E(a,b)}$ only for reals $a$ and $b$. Now we perform the
change of variables in the inner integrals over $\alpha_1$ and
$\alpha_2$:
\begin{eqnarray}
\alpha_1=\frac{\alpha + \zeta
\sqrt{2\tilde{\beta}i(\zeta_2-\zeta_1)}}{2\pi},\qquad
\alpha_2=\frac{\alpha - \zeta
\sqrt{2\tilde{\beta}i(\zeta_2-\zeta_1)}}{2\pi},
\end{eqnarray}
where here and below we assume the following branches of the square
root analytical function
\begin{eqnarray}
\sqrt{2\tilde{\beta}i(\zeta_2-\zeta_1)}=e^{i\frac{\pi}{4}\mathrm{sign}(\zeta_2-\zeta_1)}\sqrt{2
\tilde{\beta} |\zeta_2-\zeta_1|}.
\end{eqnarray}
In such a way, we arrive at the following representation that is
more convenient for the numerical calculations by the standard methods
\begin{eqnarray}\label{b2form4}
&&b^{k_1,k_2;\,k_3,k_4}_2= \int^{1}_{0} d \zeta_1 \int^{1}_0 d
\zeta_2 \min(\zeta_1,\zeta_2) \int^{\infty}_{-\infty}  \frac{d
\alpha}{2 \pi} \int^{\infty}_{-\infty} d  \zeta
\frac{e^{-\zeta^2}}{\sqrt{\pi}} \times \nonumber \\&&
\sum^{M}_{r=-M} E\left(2\pi r+\alpha+\zeta
\sqrt{2\tilde{\beta}i(\zeta_2-\zeta_1)},2\tilde{\beta}\zeta_1\right)
\tilde{E}\left(2\pi r+\alpha-\zeta
\sqrt{2\tilde{\beta}i(\zeta_2-\zeta_1)},2\tilde{\beta}\zeta_2\right)
\times \nonumber \\&& E\left(2\pi k_1+\alpha-\zeta
\sqrt{2\tilde{\beta}i(\zeta_2-\zeta_1)},2\tilde{\beta}\zeta_2\right)
E\left(2\pi k_2+\alpha-\zeta
\sqrt{2\tilde{\beta}i(\zeta_2-\zeta_1)},2\tilde{\beta}\zeta_2\right)\times
\nonumber \\&& \tilde{E}\left(2\pi k_3+\alpha+\zeta
\sqrt{2\tilde{\beta}i(\zeta_2-\zeta_1)},2\tilde{\beta}\zeta_1\right)
\tilde{E}\left(2\pi k_4+\alpha+\zeta
\sqrt{2\tilde{\beta}i(\zeta_2-\zeta_1)},2\tilde{\beta}\zeta_1\right).
\end{eqnarray}
The further numerical calculation of  \eqref{b2form4} is based ont he approximation \eqref{Eabapprox} for the functions $E$ and $\tilde{E}$.

\subsection{Zero dispersion limit}
For the case of zero dispersion $\tilde{\beta}=0$ one has the
following results:
\begin{eqnarray}\label{a1zerobeta2}
&&a_1^{n,m;p,k}\Big|_{\tilde{\beta}=0}=\int\frac{dt}{T_0}s^{(n)}(t)s^{(m)}(t)s^{(p)}(t)s^{(k)}(t)=\int^{\infty}_{-\infty}
d\tau  \mathrm{sinc}(\pi(\tau+n)) \mathrm{sinc}(\pi(\tau+m))
\mathrm{sinc}(\pi(\tau+k)) \mathrm{sinc}(\pi(\tau+p))=\nonumber \\&&
\mathrm{sinc}(\pi(n-p))\mathrm{sinc}(\pi(m-p))\mathrm{sinc}(\pi(k-p))+\frac{1}{2\pi
(k-p)}\Big(\mathrm{sinc}(\pi(n-p))\frac{\cos(\pi(k+m-2p))}{\pi(m-p)}-\nonumber
\\&&  \mathrm{sinc}(\pi(n-m))\frac{\cos(\pi(k-p))}{\pi(m-p)}-
\mathrm{sinc}(\pi(n-k))\frac{\cos(\pi(m-p))}{\pi(m-k)}+\mathrm{sinc}(\pi(n-m))\frac{\cos(\pi(k-p))}{\pi(m-k)}\Big),
\end{eqnarray}

Using the second representation in
Eq~(\ref{Eabfunction}) it is easy to obtain the simple (for
numerical calculation) result for the nondispersive channel from Eq.~(\ref{A2form4}):
\begin{eqnarray}\label{A2form4betazero}
&&A^{m_1,m_2,m_3;\,{m_4},{m_5},{m_6}}_2|_{\tilde{\beta}=0}=\frac{1}{2}
\int^{\infty}_{-\infty} d \alpha \prod^6_{i=1}
\mathrm{sinc}\left(\pi(\alpha+m_i)\right),
\end{eqnarray}

For the zero dispersion case it is easy to obtain the following representations
\begin{eqnarray}
b_1^{n,m;p,k}\Big|_{\tilde{\beta}=0}=\frac{1}{2}a_1^{n,m;p,k}\Big|_{\tilde{\beta}=0},
\end{eqnarray}
where the analytical result for $a_1^{n,m;p,k}$ is given by
Eq.~\eqref{a1zerobeta2}.

\begin{eqnarray}\label{b1zerobeta2}
b_1^{n,m;p,k}\Big|_{\tilde{\beta}=0}=\frac{1}{2}a_1^{n,m;p,k}\Big|_{\tilde{\beta}=0},
\end{eqnarray}

For zero dispersion
$\tilde{\beta}$ it is easy to obtain the one-fold integral
representation
\begin{eqnarray}\label{b2zerobeta2}
&&b^{k_1,k_2;\,k_3,k_4}_2\Big|_{\tilde{\beta}=0}=\frac{2}{3}\sum^M_{r=-M}A^{k_1,k_2,r;k_3,k_4,r}_2|_{\tilde{\beta}=0}=\frac{1}{3}\int\frac{dt}{T_0}
s^{(k_1)}(t)s^{(k_2)}(t){s}^{(k_3)}(t) {s}^{(k_4)}(t)
\sum_{r=-M}^{M}|s^{(r)}(t)|^2=\\ && \nonumber
\frac{1}{3}\int^{\infty}_{-\infty} d\tau
\mathrm{sinc}(\pi(\tau+k_1)) \mathrm{sinc}(\pi(\tau+k_2))
\mathrm{sinc}(\pi(\tau+k_3)) \mathrm{sinc}(\pi(\tau+k_4))
\sum_{r=-M}^{M}\mathrm{sinc}^2\left(\pi(\tau+r)\right),
\end{eqnarray}
where $A^{k_1,k_2,r;k_3,k_4,r}_2|_{\tilde{\beta}=0}$ is taken from
the Eq.~\eqref{A2form4betazero}.

These formulae lead to the following representations:
\begin{eqnarray}
&&2J^{r,s;{r},{s}}_{\Lambda}=4 b^{m,p;{r},{r'}}_1
b^{r',r;{m},{p}}_1-4 b^{m,p;{m},{p}}_2= \int^{+\infty}_{-\infty} d
\tau_1 \int^{+\infty}_{-\infty} d \tau_2 S^4(\tau_1,\tau_2)
-\frac{4}{3}\int^{+\infty}_{-\infty} d \tau S^6(\tau,\tau),
\end{eqnarray}
\begin{eqnarray}
&&2J^{r,s;{r},{s}}= 16b^{n,p;p,m}_1 b^{n,s;s,m}_1 +8
b^{m,p;{r},{r'}}_1 b^{r',r;{m},{p}}_1-18 b^{m,p;{m},{p}}_2=\nonumber
\\&& \int^{+\infty}_{-\infty} d \tau_1 \int^{+\infty}_{-\infty} d
\tau_2 \left(2
S^8(\tau_1,\tau_2)+4S^4(\tau_1,\tau_2)S^2(\tau_1,\tau_1)S^2(\tau_2,\tau_2)
\right) -6\int^{+\infty}_{-\infty} d \tau S^6(\tau,\tau),
\end{eqnarray}
where
\begin{eqnarray}
S^2(\tau_1,\tau_2)=\sum^{M}_{r=-M}\mathrm{sinc}(\pi(\tau_1+r))\mathrm{sinc}(\pi(\tau_2+r)),
\qquad S^2(\tau,\tau)=\sum^{M}_{r=-M}\mathrm{sinc}^2(\pi(\tau+r)).
\end{eqnarray}
And in the sum one has
\begin{eqnarray}\label{IPbetazero2}
&&2J^{r,s;{r},{s}}_{\Lambda}+2J^{r,s;{r},{s}}= 12 b^{m,p;{r},{r'}}_1
b^{r',r;{m},{p}}_1+16b^{n,p;p,m}_1 b^{n,s;s,m}_1-22
b^{m,p;{m},{p}}_2=\nonumber \\&& \int^{+\infty}_{-\infty} d \tau_1
\int^{+\infty}_{-\infty} d \tau_2 \left(3
S^8(\tau_1,\tau_2)+4S^4(\tau_1,\tau_2)S^2(\tau_1,\tau_1)S^2(\tau_2,\tau_2)
\right) -\frac{22}{3}\int^{+\infty}_{-\infty} d \tau S^6(\tau,\tau).
\end{eqnarray}
There is no necessity to calculate the two-fold integral, since it
represents the sum $12 b^{m,p;{r},{r'}}_1
b^{r',r;{m},{p}}_1+16b^{n,p;p,m}_1 b^{n,s;s,m}_1$, and for all
$b_1=\frac{1}{2}a_1$ we have the explicit representation
\eqref{a1zerobeta2}. However the formula  \eqref{IPbetazero2} is
useful to understand how  the expression in the r.h.s. of
\eqref{IPbetazero2} turns into $-(2M+1)\frac{22N_6-21N^2_4}{3}$ for
the case of the non-overlapping signals, see Eq. \eqref{IPzero}.


\end{document}